\documentclass[prc,aps,showpacs,showkeys,superscriptaddress]{revtex4}
\usepackage{amsmath}
\usepackage{amsfonts}
\usepackage{amssymb}
\usepackage{yfonts}
\usepackage{graphicx}
\usepackage{dcolumn}  
\begin{document}

\title{Continuum Shell Model}
\author{Alexander Volya}
\affiliation{Department of Physics,
Florida State University, Tallahassee, FL 32306-4350, USA}
\author{Vladimir Zelevinsky}
\affiliation{NSCL and Department of Physics and Astronomy,
Michigan State University, East Lansing, MI 48824-1321, USA}
\date{\today}

\begin{abstract}
The Continuum Shell Model is an old but recently revived method
that traverses the boundary between nuclear many-body structure
and nuclear reactions. The method is based on the non-Hermitian
energy-dependent effective Hamiltonian. The formalism,
interpretation of solutions and practical implementation of
calculations are discussed in detail. The results of the
traditional shell model are fully reproduced for bound states;
resonance parameters and cross section calculations are presented
for decaying states. Particular attention is given to one- and
two-nucleon reaction channels including sequential and direct
two-body decay modes. New calculations of reaction cross sections
and comparisons with experimental data for helium and oxygen
isotope chains are presented.
\end{abstract}
\maketitle

\section{Introduction}

New horizons are in sight in the field of nuclear physics as we
move away from the line of nuclear stability. In the everyday
experience we observe only a tiny fraction of the nuclear world,
while the recent advances in observational techniques reveal a
hidden realm of extraordinary nuclear complexes. The term
``exotic'' is commonly used to highlight the unusual nature of
newly discovered nuclear systems where structure and stability are
governed by intricate interplay of quantum many-body structure and
dynamics of nuclear reactions. Weakly bound nuclei and unstable
resonances appear as important links in the chain of nuclear
evolution in cosmos, their structure and properties are central
for energy generation in stars and production of elements in the
universe. Furthermore, the quantum objects of mesoscopic nature
are common to many fields of science including but not limited to
atoms and molecules, nanoscale condensed matter systems, atomic
clusters, atoms in traps, and prototypes of quantum
computers~\cite{WNMP04}. In many applications to mesoscopic
systems one needs to understand and utilize the features of
marginal stability and strong coupling between discrete structure
and continuum.

Mean field along with corresponding shell structure is a starting
point in the theoretical analysis of a quantum many-body system.
The shell structure itself becomes exotic on the borderline of the
continuum \cite{hansen01,brown01,caurier05,ingo05}.
Single-particle, pairwise and cluster excursions into the
continuum become essential forming halo states and resulting in
complex mixing of internal many-body states and continuum
configurations. The problems of continuum shell model (CSM), or
combining the description of reactions with the structure
calculations, have been discussed almost since the dawn of the
shell model, and summarized in the classical text \cite{mahaux69}.
However, only recently, fuelled by the discoveries of exotic
systems, acute need in theoretical understanding, and growing
computational capabilities, this subject received due attention,
so that significant advances have been made during last years. In
this work we concentrate on the version of the CSM
\cite{volya_PRC67,volya_PRL} that is going back to the Feshbach
projection formalism
\cite{feshbach58,feshbach62,rotter91,rotter01} and, even much
earlier, to the approach by Weisskopf and Wigner
\cite{weisskopf30} and works in atomic physics by Rice
\cite{rice33} and Fano \cite{fano35,fano61}. Alternative
formulations, such as \cite{betan02,michel02,michel03,michel04},
see a review \cite{okolowicz03}, were also successfully developed
recently.

The specific attractive features of the approach discussed below
are the natural unification of structure and reactions, full
agreement with the results of the traditional shell model (SM) in
the discrete spectrum, correct energy behavior of resonance widths
and reaction cross sections near thresholds, self-consistent
consideration of isotope chains, and exact unitarity of the
scattering matrix. At this point we use the standard residual
interactions adjusted in numerous applications of the conventional
SM, although the problem of better interactions in the continuum
remains open (the first steps in this direction were made in Ref.
\cite{hagen05}).

We organize the discussion here starting with the formal
description of the CSM approach in the following section; this
will be the formalism that we imply under the term CSM throughout
this work. In Sec.~\ref{sec::csm} we also discuss mathematical
details of the formulation, relation to observables and the
unitarity  of the scattering matrix. Sec.~\ref{sec::III} is
devoted to the detailed consideration of different parts of the
interaction; we consider the limit of the conventional shell model
one-body decay channels, and sequential and direct two-body
channels. The realistic applications are shown and compared with
experiment in Sec.~\ref{sec::appl}.

\section{Formulation of the Continuum Shell Model\label{sec::csm}}

\subsection{Effective Hamiltonian}

In what follows we assume that the many-body Hilbert space is
spanned by Slater determinants constructed from the
single-particle (s.p.) orbitals $|j\rangle$ in the mean field.
Using the notations of secondary quantization we denote the s.p.
creation and annihilation operators for discrete orbitals as
$b^\dagger_j$ and $b_j$ labelled by a combined discrete label $j$.
For the continuum states we use s.p. operators
$b^\dagger_j(\epsilon)$ and $b_j(\epsilon)$ which are labelled
with the discrete index $j$ and the continuous s.p. energy
variable $\epsilon$. If properly constructed for a mean-field
potential, see below the discussion related to one-body reactions,
these states are automatically orthogonal and form a complete set.
In this work, however, we require only separate orthogonality of
the bound states, $[b_j,b^\dagger_{j'}]_{+}=\delta_{jj'}$, and of
the continuum states normalized according to
$[b_j(\epsilon),b^\dagger_{j'}(\epsilon')]_{+}=
\delta(\epsilon-\epsilon')\,\delta_{jj'}.$

The full many-body space can be separated into two parts. The set
of $N$-particle bound states $|1;N\rangle=b^\dagger_{j_1}\dots
b^\dagger_{j_N}|0\rangle$ forms the ``internal'' space, ${\cal
P}$, where the index $1=\{j_1,j_2\dots j_N\}$ labels the Slater
determinant, the $m$-scheme representation in the SM terminology.
The remaining space, ${\cal Q}$, is the ``external'' continuum,
i.e. many-body states which contain one or more continuum s.p.
orbitals. The external many-body states  $|c,E\rangle$ are
labelled by total continuum energy $E$ and the set of asymptotic
variables $c$ that defines a reaction channel (it includes the
characteristics of the residual nucleus). The channel variable $c$
is discrete only in the case of the one-body decay where energy
conservation fully determines energies of the two decay products.
In general this variable is continuous, containing relative
energies or momenta of decay products needed for full
specification of the final state. In the description of the
formalism below we use notations $\sum_c$ and $\delta_{c c'}$
which in the case of the continuous channel variables should be
interpreted as $\int dc$ and $\delta(c-c')$, respectively. In Sec.
\ref{sec::2n}, where channel $c$ in the two-body decay implies the
presence of the relative energy variable, the sum over channels is
explicitly given in terms of an integral. By construction, the
many-body states in ${\cal P}$- and ${\cal Q}$- spaces are
mutually orthogonal and normalized as
\begin{equation}
\langle 1|2\rangle =\delta_{1 2} \,,\quad \langle c;E|c';E'
\rangle =\delta_{c c'} \delta(E-E').           \label{0}
\end{equation}

Within the total space ${\cal P}+{\cal Q}$, we solve the
stationary Schr\"odinger equation
\begin{equation}
H|\alpha;E\rangle =  E|\alpha;E \rangle,           \label{1}
\end{equation}
where the full wave function $|\alpha;E\rangle$ is in general a
superposition of internal states $|1\rangle$ and external states
$|c;E'\rangle$,
\begin{equation}
|\alpha;E \rangle = \sum_1 \alpha_1(E) |1 \rangle + \sum_c \int
dE'\,\alpha_c(E';E)|c; E'\rangle.                   \label{2}
\end{equation}
The Hamiltonian $H$ has parts acting within and across ${\cal P}$
and ${\cal Q}$ spaces, $H=H_{\cal PP}+H_{\cal PQ}+H_{QQ}.$ The
external states can be eliminated by introducing a propagator that
acts exclusively within ${\cal Q}$ space,
\begin{equation}
G_{\cal QQ}(E)=\frac{1}{E-H_{\cal QQ}+i0}, \label{propag}
\end{equation}
where the infinitesimal imaginary displacement selects the
appropriate boundary conditions for the scattering problem. Next
we assume that the channel labels $c$ correspond to the
eigenchannels \cite{engelbrecht73} in the ${\cal Q}$ space,
$H_{\cal QQ}|c; E\rangle =E |c;E\rangle $. Then the Schr\"odinger
equation (\ref{1}) ``projected'' into the subspace $P$ becomes
\begin{equation}
\sum_2 \left [\langle 1|H|2 \rangle + \sum_c \int dE'
\frac{\langle 1|H-E|c;E'\rangle \langle c;E'|H-E|2\rangle }
{E-E'+i0} -\delta_{1 2} E \right ] \alpha_2 =0. \label{incoef}
\end{equation}
The amplitude of the continuum admixture in the full wave function
(\ref{2}) is
\begin{equation}
\alpha_c(E';E)=\frac{\sum_1 \alpha_1(E) {A^c_1}^*(E',E)}{E-E'+i0},
                                               \label{chi}
\end{equation}
where we introduced notations for the $H_{{\cal PQ}}$ coupling
amplitude
\begin{equation}
A_1^c(E',E)=\langle 1|H-E|c;E' \rangle.        \label{ampa}
\end{equation}
This amplitude depends on the continuum variable of energy $E'$
and running energy $E$. As follows from the definition
(\ref{ampa}), there is no explicit $E$-dependence when internal
and external spaces are orthogonal. However, the important
$E'$-dependence remains; the kinematic factors included in the
definition of the channel states $|c;E'\rangle$ ensure that the
phase space shrinks to zero and the channel $c$ becomes closed
below threshold energy $E_{c}$ characteristic for a given channel.
Whence, the amplitudes $A^{c}_{1}$ vanish at $E'<E_{c}$.

The set of equations (\ref{incoef}) for coefficients $\alpha_1$
looks as an eigenvalue problem with the effective Hamiltonian
matrix ${\cal H}$ in the intrinsic space defined as
\begin{equation}
\langle 1 |{\cal H}({ E})|2 \rangle = \langle 1|H|2 \rangle +
\sum_c \int dE' \frac{A_1^c(E',E)\,A_2^{c\ast}(E',E)}{E-E'+i0}.
                                               \label{heff1}
\end{equation}
The integral in Eq. (\ref{heff1}) can be further decomposed into
its Hermitian part (principal value), $\Delta(E)$, and the
anti-Hermitian part, $-(i/2)W(E)$,
\begin{equation}
\sum_c \int dE'\,\frac{A^c_1(E',E)A^{c\ast}_2(E',E)}{E-E'+i0}=
\sum_{c} \,{\rm P.v.} \int dE' \,\frac{A^c_1(E',E)A^{c\ast}_2
(E',E)}{E-E'} - i \pi \sum_{c\,({\rm open})}
A^c_1(E)A^{c\ast}_2(E)\,,                     \label{6}
\end{equation}
where $A^{c}_{1}(E)\equiv A^{c}_{1}(E,E)$. Thus, the effective
Hamiltonian for the ${\cal P}$ space (\ref{heff}) takes form
\begin{equation}
{\cal H}(E)=H_{\cal PP}+\Delta (E) -\frac{i}{2}\,W (E)\,,
                                               \label{heff}
\end{equation}

\begin{figure}[ht]
\vskip -0.4 cm
\includegraphics[width=9 cm]{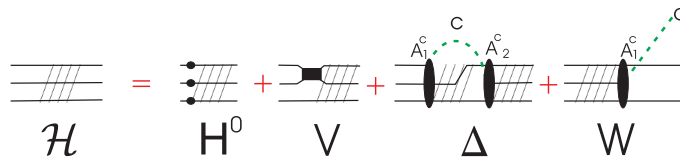}
\caption{(Color online) Diagrammatic equation for the full
Hamiltonian corresponding to the dynamics in the ${\cal P}$ space,
Eq. (\ref{heff}). Parts $H^\circ$ and $V$ on the figure indicate
the one- (s.p. energies) and two-body parts of the internal
Hamiltonian $H_{\cal PP}$.\label{fig::nhh}}
\end{figure}

The resulting dynamics generated by the effective Hamiltonian
(\ref{heff}) contains a usual ${\cal P}$-space contribution that
we identify here with the traditional SM corrected by the virtual
``off-shell'' excitations into the continuum  via the self-energy
term,
\begin{equation}
\langle 1 |\Delta (E)|2\rangle = \sum_{c} \,{\rm P.v.} \int dE' \,
\frac{A^c_1(E',E)\, A^{c\ast}_2(E',E)}{E-E'},  \label{8}
\end{equation}
and supplemented by the anti-Hermitian term,
\begin{equation}
\langle 1 | W(E)|2\rangle = 2\pi  \sum_{c\,({\rm open})}
A^c_1(E)\,A^{c\ast}_2 (E),                   \label{weff}
\end{equation}
that represents the irreversible departure into ${\cal Q}$-space,
i.e. decays; see Fig.~\ref{fig::nhh}. The term $W$ that comes from
the poles in integration (\ref{heff1}), has a factorized form,
which is shown below to relate to conservation of probability and
unitarity of the scattering matrix. The amplitudes $A^c(E)$ here
represent on-shell processes that depend only on one energy
parameter and correspond to real decays with energy conservation,
$E'=E$. These poles in integration appear only when running energy
$E$ is above the decay threshold $E_{c}$ for a channel $c$. The
channels where decays are allowed are referred to as open. At
$E\rightarrow E_{c}+0$ the amplitudes $A_{1}^{c}$ vanish due to
the kinematic factors implicitly included into their definition.

\subsection{Scattering matrix and unitarity\label{cross}}

The ``outside"  view from the reaction side of the problem is
equally important. In accordance with general scattering theory,
the transition matrix,
\begin{equation}
T^{cc'}(E) = \sum_{12}A^{c\ast}_{1}(E)\, \left(\frac{1}{E-{\cal
H}(E)}\right)_{12}A^{c'}_{2}(E), \label{tmat}
\end{equation}
describes the process that starts in the entrance channel $c'$
with amplitude $A^{c'}_2$ originated from the interaction $H_{\cal
PQ}$, continues through internal propagation within the ${\cal
P}$-space driven by the non-Hermitian energy-dependent effective
Hamiltonian (\ref{heff}) (with all excursions into ${\cal Q}$
space included), and ends by exit to the channel $c$ described by
the amplitude $A^{c\ast}_{1}(E)$. The scattering matrix can be
written as
\begin{equation}
S^{cc'}(E)=\exp(i\xi_c)\,\left\{\delta^{cc'}-2\pi
i\,T^{cc'}(E)\right\} \exp(i\xi_{c'}).
\label{scatt}
\end{equation}
The additional phase shifts $\xi_{c}(E)$ describe the potential
scattering or a contribution of remote resonances outside of the
valence space of the model.

The factorized nature of the non-Hermitian contribution to the
effective Hamiltonian is the key for conserving the unitarity of
the $S$ matrix \cite{durand76}. This can be demonstrated by
considering the propagator for the effective Hamiltonian
\begin{equation}
{\cal G}(E)=\frac{1}{E-{\cal H}}                  \label{calG}
\end{equation}
generated from the unperturbed propagator for the full Hermitian
part,
\begin{equation}
G(E)=\frac{1}{E-H_{\cal PP}-\Delta(E)}.        \label{G}
\end{equation}
With $W=2\pi {\bf A} {\bf A}^\dagger$, where ${\bf A}$ represents
a channel matrix (a set of columns of vectors $A^c_1$ for each
channel $c$), we iterate the Dyson equation
\begin{equation}
{\cal G}=G-(i/2)G W{\cal G}                \label{Dyson}
\end{equation}
and, due to the factorized form of $W$, come to
\begin{equation}
{\cal G}=G-i \pi  G {\bf A}\,\frac{1}{1+i \pi {\bf A}^\dagger G
{\bf A}}\, {\bf A}^\dagger G,            \label{woodbury}
\end{equation}
that is called the Woodbury equation in mathematical literature.

The transition matrix, $T={\bf A}^\dagger {\cal G} {\bf A}$, can
then be written with the aid of the matrix $R={\bf A}^\dagger G
{\bf A}$ that is analogous to the $R$-matrix of standard reaction
theory. The unitarity of $S$-matrix follows directly from these
equations, see also \cite{sokolov89},
\begin{equation}
T=\frac{R}{1+i\pi R}\,,\quad S=\frac{1-i\pi R}{1+i\pi R}.
                                            \label{trmat}
\end{equation}

\subsection{Energy dependence and resonances}

The effective Hamiltonian of Eq. (\ref{heff}) is energy-dependent:
at each scattering energy $E$ its running eigenvalues are complex
numbers ${\cal E}_\alpha(E)$. This highlights the structure in Eq.
(\ref{2}) that the eigenstate is a superposition of internal
states and asymptotic decay states that have right energy. The
relatively small, and numerically tractable, dimension of basis
states, in exchange for non-Hermicity and energy dependence, is a
noteworthy advantage of this method as compared to direct
discretization of continuum used in other approaches.

The eigenvalue problem involving a complex matrix of a general
form requires finding two sets of adjoint eigenvectors: left,
$|\alpha \rangle$, and right, $|\tilde{\alpha}\rangle$. They
satisfy
\begin{equation}
{\cal H} |\alpha \rangle = {\cal E}_\alpha |\alpha \rangle \quad
{\rm and} \quad\, \langle \tilde{\alpha} |{\cal H} = {\cal
E}_\alpha^* \langle \tilde{\alpha} |.        \label{intro::eigen}
\end{equation}
The left and
right eigenstates correspond to time reversed motions; they no
longer have to coincide because the ${\cal T}$-invariance in the
internal space is broken by irreversible decays. The global
symmetry with respect to the direction of time is however
maintained by the full Hamiltonian that includes the products of
reactions. As a result, the left and right eigenstates have the
wave functions interrelated by complex conjugation which is the
time inversion operation. The Hermitian conjugation of the
Hamiltonian switches the roles of left and right; the same effect
can be reproduced by selecting an advanced propagator boundary
condition in (\ref{propag}) discussed earlier. The biorthogonality
relation is given by
$\langle\tilde{\alpha}|\beta\rangle=\delta_{\alpha\beta}$;
similarly, the expectation value of an operator $X$ is $\langle
\tilde{\alpha}|X|\beta \rangle$. These properties, the formalism
of CSM and its interpretation become transparent in the one-body
problem discussed below in Sec.~\ref{sec::mf}; for further notes
on this topic we refer to \cite{berggren68}.

There are various interpretations of eigenvalues ${\cal
E}_{\alpha}(E)$ of Eq. (\ref{incoef}). Only for bound states below
all decay thresholds a condition ${\cal E}_\alpha(E)=E$ is
satisfied since $W=0.$ With non-zero $W$, the eigenvalues of the
effective Hamiltonian are in general complex,
\begin{equation}
{\cal E}_\alpha(E)=E_{\alpha}(E)-\frac{i}{2}\,\Gamma_{\alpha}(E),
                                          \label{reson}
\end{equation}
describing the quasistationary states. These resonances and their
widths $\Gamma_{\alpha}\geq 0$ satisfy the Bell-Steinberger
relation
\begin{equation}
\langle\tilde{\alpha}|W|\alpha\rangle=\Gamma_\alpha, \label{BS}
\end{equation}
where the left hand side can be expressed through the amplitudes
$A^{c}_{1}$ transformed to the biorthogonal basis of
quasistationary states $|\alpha\rangle$.

The continuation of the original problem to the lower part of the
complex energy plane, $E\rightarrow{\cal E}=E-(i/2)\Gamma$ allows the
condition
\begin{equation}
{\cal E}_\alpha({\cal E})={\cal E}.           \label{scond}
\end{equation}
The complex energy roots here can be identified with the poles in
the scattering matrix (\ref{scatt}), and the states correspond to
many-body resonant Siegert states \cite{siegert39} since by
construction the eigenstate (\ref{2}) is a regular function with
outgoing asymptotics. One can also use the Breit-Wigner approach
\cite{breit36} and identify resonances differently, with a
condition
\begin{equation}
{\rm Re}[{\cal E}_\alpha(E)]=E,\quad \Gamma_{\alpha}=-2{\rm Im}
[{\cal E}(E)].                                  \label{bwcond}
\end{equation}
In the limit of small imaginary part (narrow resonances), both
definitions are equivalent. However in general the difficulty in
parameterizing the resonance width and centroid energy is related
to the non-exponential character of decay caused by the energy
dependence of the Hamiltonian parameters. Wide resonances cover
broad regions of energy and therefore are particularly affected by
this dependence. This leads to the  non-generic and asymmetric
shape of the resonance cross section that makes standard
Breit-Wigner or Gaussian parameterizations inappropriate.

With either definition, Eqs. (\ref{scond}) or (\ref{bwcond}) for
resonant states are complicated sets of nonlinear equations. In
some cases \cite{volya_PRC67} the condition (\ref{scond}) may lead
to unphysical solutions. For the realistic calculations shown in
Sec.~\ref{sec::appl} we select the Breigt-Wigner definition
(\ref{bwcond}) and implement an iterative approach starting from
the energy determined by the conventional SM with $W=0$. Although
it is convenient to express the solutions in terms of resonant
states, the parameters are definition-dependent and become
misleading for broad states or in the case of overlapping
resonances when interference is important. In these cases one
should turn to the observable scattering cross section determined
by the $S$-matrix of Eq. (\ref{scatt}). The computation of the
scattering cross section is a problem of matrix inversion which is
linear in accordance with physical principles, but it has to be
done at each energy which can make this task numerically unstable
for narrow resonances. The cross sections can be calculated using
the $R$-matrix of Eq. (\ref{trmat}) and the Woodbury equation in
which case complex arithmetics can be avoided. We will see
complementary pictures that can be obtained by calculating the
cross section and via resonance parameters coming from the
diagonalization of the effective Hamiltonian in the example shown
in Sec.~\ref{sec::appl}.

\section{From Hamiltonian to dynamics\label{sec::III}}

The derivation above is based on the decomposition of the full
Hamiltonian into $H=H_{\cal PP}+H_{\cal PQ}+H_{QQ}$. Another
useful classification traditional to the SM is by the types of
many-body processes it can generate. At this stage we restrict the
interactions by one- and two-body and limit the space ${\cal Q}$
by the states with only one or two nucleons in continuum. The
typical shell model limitations by few valence shells are imposed
on the intrinsic  space ${\cal P}$. In this framework we define
the full Hamiltonian and discuss processes associated with each of
the terms.

Our discussion here, however, does not touch the lack of knowledge
of the effective interaction. Although sophisticated methods of
deriving the effective interactions were suggested
\cite{sn132g,hagen05}, the best results and the most predictive
power in the conventional SM come from the phenomenological
interactions, such as USD \cite{USD}, fitted to experimental data.
The situation becomes increasingly more complicated when effective
interactions involving continuum are to be used \cite{berggren68}.
In this work we identify the internal
interaction $H_{\cal PP}$, together with the Hermitian self-energy
term $\Delta$ included, with a shell model Hamiltonian of the form
\begin{equation}
H_{\cal PP}+\Delta=\sum_j \epsilon_j b^\dagger_j b_j + \frac{1}{4}
\sum V(j_1 j_2; j_3 j_4) b^\dagger_{j_1} b^\dagger_{j_2} b_{j_3}
b_{j_4}.
\label{smv}
\end{equation}
The parameters, s.p. energy levels and antisymmetrized two-body
matrix elements, are known from fits to experimental data, such as
\cite{USD}, see also review \cite{brown01}, and available from
interaction libraries, such as \cite{oxbash}. The energy
dependence of these parameters that comes from $\Delta(E)$ is
ignored here since it has not been included in the fitting
process. As demonstrated below this dependence is weak and smooth
function of energy. Defining the internal Hamiltonian in this way
we guarantee that below thresholds the CSM provides the results
identical to the well established SM with effective interactions.
Above thresholds, the shell model interactions were fitted to
experimental data using $R$-matrix analysis that identifies
interactions (\ref{smv}) in the same way, Eq.~(\ref{trmat}). The
consistent readjustment of shell model interaction parameters
taking into account energy dependence of Hermitian part is beyond
the scope of this paper but remains a subject of future work.

In this work we assume that the Hamiltonian describing the motion
of nucleons in the continuum is purely single-particle
\begin{equation}
H_{\cal QQ}=\sum_j\int d\epsilon\, \epsilon
\,b^\dagger_j(\epsilon)b_j(\epsilon).             \label{hqq}
\end{equation}
The asymptotic one- and two-particle continuum states considered
in this work are antisymmetrized products of internal eigenstates
$\alpha$ of the residual nucleus and the wave function(s) of
particle(s) in the continuum. The states in one-body channels,
\begin{equation}
| c,E \rangle = b^\dagger_j(\epsilon_j) |\alpha;N-1\rangle \,,
\quad E=E_\alpha+\epsilon_j, \label{cwf}
\end{equation}
are labelled by energy $E$ and the discrete channel index $c$ that
combines the s.p. quantum numbers $j$ and characteristics $\alpha$
for the eigenstate of an $(N-1)$-particle daughter system.

The assumption (\ref{hqq}) allows one to express similarly the
two-nucleon channel states,
\begin{equation}
|c;E\rangle = b_j^\dagger(\epsilon) b^\dagger_{j'}(\epsilon')
|\alpha;N-2 \rangle, \label{twob}
\end{equation}
characterized by the total energy
$E=E_\alpha+\epsilon_j+\epsilon_{j'}$ combined of the daughter
binding energy $E_\alpha$ and energies of emitted nucleons in s.p.
states $j$ and $j'$. Here the channel $c$ contains in addition
information on continuous relative energy distribution between
emitted particles. Within this work we do not consider cases when
both particles in the continuum are charged thus the assumptions
in Eqs. (\ref{hqq}) and (\ref{twob}) are sufficient. The
generalization for bound states of two particles in the continuum
is relegated to the next stage.

\subsection{Single-particle decay and resonances \label{sec::mf}}

Now we need to define the interaction $H_{\cal PQ}$ responsible
for coupling of intrinsic space with the continuum. We start with
the one-body part and associate it with some potential $V$
assuming this potential to be spherically symmetric.

We first treat a pure s.p. problem of a particle moving in the
mean field or independent particle shell model. The generalization
covering all s.p. channels in many-body cases is straightforward
and discussed in what follows. Of course, this subject is
extensively covered by textbooks \cite{BMI,landauIII,merzbacher}.
The purpose of the formulation presented here is to emphasize the
conceptual identity between the full CSM and its trivialized
version represented by a single particle in a potential. The
notions and definitions of resonances, scattering matrix and its
poles, time reversal and non-Hermiticity already appear in this
simplest case. This section also highlights some technical details
used later for the s.p. part in the full CSM including the generic
threshold behavior of the decay amplitudes.

In the coordinate representation the Schr\"odinger equation for
the radial part of the s.p. wave function,
\begin{equation}
\langle {\bf r}|b^\dagger_j|0\rangle=[Y_{l} \chi]_j
\frac{u_j(r)}{r},
\end{equation}
where $Y_{l}$ and $\chi$ represent the angular and spin parts
coupled to total angular momentum $j$, is
\begin{equation}
\left \{-\frac{d^2}{dr^2}+\frac{l(l+1)}{r^2}+ 2\mu\left[V(r)+e^2\,
\frac{Z z}{r}\right] \right \} u_j(r) =k^2 u_j(r), \label{gf:1}
\end{equation}
where $k^2=2\mu \epsilon$, $\mu$ is the reduced mass, $z$ and $Z$
are charges of the particle and of the residual nucleus,
respectively. The spin-orbit part can be included here assuming
that the potential $V(r)$ depends on $l$ and the spin orientation
which in our notations are hidden in the s.p. index $j$.

For Eq. (\ref{gf:1}) with $V(r)=0$, we have the regular, $F_l(k
r),\,F_l(0)=0,$ and the irregular, $G_l(k r)$, solutions as
Coulomb wave functions with the charge parameter $\eta=\mu \alpha
Z z/k$. For a neutral particle, $z=0$, the regular and irregular
solutions can be expressed in terms of spherical Bessel and
Neumann functions,
\begin{equation}
F_l(kr)=kr j_l(kr)\,,\quad G_l(kr)= -kr n_l(kr).   \label{bessel}
\end{equation}
The two independent solutions are related by the Wronskian,
\begin{equation}
G_l \frac{d}{d (kr)} F_l -F_l \frac{d}{d (kr)} G_l =1.
                                            \label{wronskian}
\end{equation}
Thus, the ${\cal Q}$ space states are energy-normalized regular
solutions,
\begin{equation}
\langle {\bf r}|j;\epsilon \rangle =\langle
{\bf r}|b^\dagger_j(\epsilon)|0\rangle=[Y_{l} \chi]_j
\sqrt{\frac{2\mu}{\pi k}}\,\frac{F_l(kr)}{r}.
\end{equation}
Following the definition in Eq. (\ref{ampa}), the s.p. decay
amplitude is
\begin{equation}
a_j(\epsilon_j,\epsilon)=\langle j|H_{\cal PQ}-
\epsilon| j;\epsilon_j\rangle=\sqrt{\frac{2 \mu}{\pi k_j}}\int_0^\infty dr\,
F_l(k_jr) \left [V(r) + \epsilon_j-\epsilon\right ]u_j(r) \,.
                                         \label{spA}
\end{equation}
A positive energy internal state $u_j(r)$ decays with the width $
\gamma_j=2\pi a_j^2$ determined by the above equation; under this
choice of phases, the decay amplitudes are real. The s.p.
amplitude has only one index $j$.

This result for the decay width can be reproduced through the
equivalent consideration of the on-shell scattering process. Let
us introduce incoming and outgoing (Coulomb) waves
$O^{\pm}_l(r)=G_l(r)\pm i F_l(r)$. Consider a resonant state
$u_j$; since through the rest of this subsection we concentrate on
a state with a given set of s.p. quantum numbers $j$ we will omit
this subscript in notations, the orbital momentum subscript $l$
which is a part of the combined index $j$ is also omitted. The
state $u$ can be normalized as a discrete state when the decaying
component that obeys the Siegert \cite{siegert39} outgoing wave
boundary condition,
\begin{equation}
\lim_{r\rightarrow\infty} u(r)={\cal N}\, O^+(kr),
\end{equation}
is neglected. Then the outgoing flux normalized by velocity
determines the decay width
\begin{equation}
\gamma=2\pi a^2= \frac{k}{\mu} |{\cal N}|^2.   \label{spwidth}
\end{equation}
It follows from here that the asymptotics of the decaying states
are given by the decay amplitude,
\begin{equation}
\lim_{r\rightarrow\infty} u(r)= -\sqrt{\frac{2\pi \mu}{k}}\,
a(\epsilon) O^+(kr),
\end{equation}
where we selected a phase to be consistent with the previous
definition. Using the Wronskian relations, the outgoing part can
be extracted from the wave function $u_j$ leading to
\begin{equation}
a(\epsilon)=-\sqrt{\frac{1}{2 \pi \mu k}} \left . \left
(u\,\frac{dF}{dr}\, - F\, \frac{du}{dr} \right )
 \right |_{r\rightarrow\infty}\,.         \label{spA1}
\end{equation}
This equation is identical to Eq. (\ref{spA}) since the
Schr\"odinger equation (\ref{gf:1}) that must be used to determine
the outgoing component guarantees that
\begin{equation}
\frac{d}{dr} \left ( u\, \frac{d F}{dr}-F\, \frac{du}{dr} \right
)=-2 \mu F V(r) u(r)\,,           \label{gf:helper}
\end{equation}
where $F$ is any of the Coulomb wave functions.

The eigenstate wave function in the asymptotics can be expressed
via the s.p. scattering phase shift,
$u_j(\epsilon)\sim\cos(\delta_j) F_l + \sin(\delta_j) G_l $. The
related $S$-matrix then can be found as
\begin{equation}
S=\exp(2i\delta)=\left . \frac{u\frac{d}{dr} O^{-} -O^{-}
\frac{d}{d r} u }{u\frac{d}{dr} O^{+} -O^{+} \frac{d}{d r} u
}\right |_{r\rightarrow \infty}              \label{1:sm}
\end{equation}
that is consistent with definitions (\ref{tmat}) and
(\ref{scatt}). The poles of the scattering matrix correspond to
the condition of the regular wave function with the outgoing wave
in asymptotics. Just as in the general case in
Sec.~\ref{sec::csm}, this can not be satisfied at real energy,
while if the problem is taken into a complex energy plane,
$k\rightarrow\kappa=k-i\varkappa$ and $\epsilon\rightarrow
e=\epsilon - i\gamma/2$, the discrete set of solutions emerges.
Thus, the resonance energy $\epsilon$ and the width $\gamma$ can
be defined as real and imaginary parts, respectively, of the
complex energy which is the pole of the scattering matrix.
Consistent with the general theory, the time reversed problem is
physically equivalent; the boundary condition then is that the
wave function be regular at the origin and represent an incoming
wave in asymptotics. This is a ${\cal T}$-reversed state, with the
corresponding ``left'' momentum eigenvalue $\tilde{\kappa}$ that
is related to that of the ``right'' eigenstate as
$\tilde{\kappa}=-\kappa^*$. This agrees with the symmetry
properties of the $S$ matrix,
\begin{equation}
S(\kappa)= S^*(-\kappa^*)=S^{-1}(-\kappa),
\end{equation}
and assures that left and right energy eigenvalues are complex
conjugate.

Numerically, the decay amplitudes can be calculated directly from
(\ref{spA}) or (\ref{spA1}); it has been demonstrated in
\cite{davids00} for proton emitters that these methods are equally
effective in practice. The effective non-Hermitian s.p.
Hamiltonian can be solved resulting in Gamow states via an
iterative procedure based on the Green's function, similar to the
approach discussed in \cite{burgov89,burgov96}. Green's function
is constructed for the free particle case $V=0$ using Coulomb
functions at some momentum $k_0$ and the Siegert boundary
conditions,
\begin{equation}
{\cal G}(r, r')=\frac{1}{k_0}\, F(k_0 r_{<}) O^+(k_0 r_{>}).
\end{equation}
The $r_{<}$ and $r_{>}$ denote the smaller and the larger of $r$
and $r'$, respectively. The integral equation for the resonant
state becomes
\begin{equation}
u(r)=\int _0^\infty {\cal G}(r,r') \left[\kappa^2-k_0^2-2\mu V(r')
\right ]u(r') dr'.                     \label{gf:gam1}
\end{equation}
This leads to the following equation for the radial part $u(r)$:
\begin{equation}
u(r)=\frac{1}{k_0} F(r) \left \{\int_r^\infty O^+(r')
\left[\kappa^2-k_0^2-2\mu V\right ]u(r')dr' \right \}+
\frac{1}{k_0} O^+(r) \left \{\int_0^r F(r')
\left[\kappa^2-k_0^2-2 \mu V\right ]u(r')dr' \right \}.
                                          \label{gf:gam2}
\end{equation}
The complex momentum $\kappa$ is determined self-consistently with
the decay flux defined by the outgoing component in the above
equation.

\subsection{Threshold behavior}

The behavior of the decay width and the self-energy term
$\Delta(\epsilon)$ in the vicinity of threshold is particularly
important. The one-body decay is an instructive example of the CSM
at work. Following the above definitions we evaluate
\begin{equation}
\int_0^\infty d\epsilon_j \frac{|a_j(\epsilon_j,
\epsilon)|^2}{\epsilon-\epsilon_j+i0}=
\Delta(\epsilon)-\frac{i}{2} \gamma(\epsilon)     \label{speq}
\end{equation}
in the vicinity of the single-particle threshold at zero energy
using Eq. (\ref{spA}) which we decompose as
\begin{equation}
a_j(\epsilon_j,\epsilon)=\langle j|V| j;\epsilon_j\rangle+
(\epsilon_j-\epsilon) \langle j| j;\epsilon_j\rangle. \label{spB}
\end{equation}
Only the first term in (\ref{spB}) leads to a pole in Eq.
(\ref{speq}). The main contribution comes from low-energy
scattering states, namely in the limit when the de Broglie
wavelength of the scattered particle exceeds the range of the
potential. For a charged particle, the energy behavior of the
amplitude (\ref{spA}) follows from that of the regular Coulomb
function. For a neutral particle, in this limit $F_l=
(kr)^{l+1}/(2l+1)!!\propto\epsilon^{(l+1)/2}$. Thus, we can assume
that
\begin{equation}
\langle j|V| j;\epsilon_j\rangle={\ae_j}\,
(\sqrt{\epsilon_j})^{l+1/2},
\end{equation}
where the constant ${\ae}_j$ is
\begin{equation}
\ae_j=\frac{(2\mu)^{(l+3/2)/2}}{\sqrt{\pi}(2l+1)!!}
\int_0^\infty r^{l+1} V(r) u_j(r) dr\,.
\end{equation}

In the integral (\ref{speq}) only the term $\propto\ae^2$ contains
the contribution from the pole if $\epsilon>0$. Direct integration
produces the result
\begin{equation}
\Delta(\epsilon)=\langle j|-H_{\cal QQ}-\hat{\cal Q}V-V\hat{\cal
Q} |j\rangle + \epsilon \langle j|\hat{\cal Q}|j\rangle+\pi \ae^2
\Theta(-\epsilon)\epsilon^l\sqrt{-\epsilon} \,, \label{spre}
\end{equation}
and for imaginary part
\begin{equation}
\gamma_j(\epsilon)=2\pi \ae_j^2 \Theta(\epsilon) \epsilon^{l+1/2}.
                                                \label{plaw}
\end{equation}
Here $\hat{\cal Q}=\int d\epsilon_j |j\epsilon_j\rangle \langle j
\epsilon_j|$ is the projection operator into ${\cal Q}$ space;
$\Theta$ is the Heaviside step function. The first two terms in
Eq. (\ref{spre}) appear as a correction to energy due to the
possible non-orthogonality between the states of ${\cal Q}$ and
${\cal P}$. They are small in any reasonably selected situation;
they are identically zero in the examples shown below where spaces
${\cal P}$ and ${\cal Q}$ are obtained from the full numerical
solution of the Woods-Saxon potential. Only the second term in
(\ref{spre}) depends linearly on energy.

The last term in Eq. (\ref{spre}) appears only below threshold and
represents virtual excitation into continuum induced by
interaction. This contribution is a continuous function of energy,
that is also smooth except for the $s$-wave neutral particle decay
channel. The behavior of the self energy $\Delta(\epsilon)$
including the $s$-wave $\sim\sqrt{\epsilon}$ type cusp is well
studied  in hadron physics, see also \cite{landauIII,baz71}. The
scaling coefficient $\ae_j$ that enters this expression is
typically small; the same coefficient determines the energy
scaling of the decay width above threshold, Eq. (\ref{plaw}), see
Fig.~\ref{weo16wss}. The behavior $\gamma\sim \epsilon^{l+1/2}$ is
consistent with the phase-space volume for one-body decay. It is
universal, so that, if the potential is being adjusted to
reproduce a certain resonance energy, then the behavior of the
width as a function of resonance energy remains the same
\cite{wigner48,baz71}.

\subsection{One-body channels in a many-body system\label{sec::ob}}

The one-body decay amplitude in a many-body system is given by the
s.p. decay and the spectator overlap,
\begin{equation}
A^c_1(E)= a_j(\epsilon)\,\, \langle 1; N|b^\dagger_j|\alpha;N-1\rangle,
\label{onebcont}
\end{equation}
where energy of the continuum state is $E=E_\alpha+\epsilon$ and
$c=\{\alpha, j\}$. This amplitude is to be directly used in the
non-Hermitian effective Hamiltonian (\ref{heff}). Even on the
level of s.p. decays, the approach outlined here goes beyond the
consideration based solely on spectroscopic factors. Here the
non-Hermitian part of the effective Hamiltonian is not a s.p.
operator. Indeed, even s.p. decays can generate significant
restructuring inside the nucleus. Effects, such as shape changes
or changes in pairing coherence, are extremely important for the
physics of nuclei far from stability. The energy dependence that
was discussed earlier is another distinct feature.

Although in all calculations presented in this work we use a
general form of the effective Hamiltonian, below we show a set of
approximations that establish a correspondence with the
traditional SM description of decay. For simplicity we assume that
all internal s.p. states can be identified by spin and parity,
i.e. the space is small enough not to include several major
shells, so that there is no need in main quantum numbers. The
non-Hermitian term in the full Hamiltonian coming from s.p. decay
channels is then diagonal,
\begin{equation}
\langle 1 | W(E)|2\rangle = 2\pi \delta_{1 2} \sum_{c\,({\rm
open})} |a_{j}(\epsilon_j)|^2 |\langle
\alpha;N-1|b_j|1;N\rangle|^2,                 \label{W1}
\end{equation}
where s.p. energy satisfies $E=\epsilon_j+E_\alpha;$ below we
suppress the energy argument if it is unambiguous.

In the case of remote thresholds, the decay amplitudes become
essentially independent of energy, and the set of continuum
channels $c=\{\alpha,j\}$ includes almost all possible daughter
states $\alpha$. The completeness in $\alpha$ and energy
independence can be then used to simplify (\ref{W1}),
\begin{equation}
\langle 1 | W|2\rangle = 2\pi \delta_{1 2} \sum_{j} |a_{j}|^2\,
\langle 1;N|b^\dagger_j b_j|1;N\rangle.       \label{W2}
\end{equation}
As a result, $W$ becomes a s.p. operator that assigns a width
(\ref{spwidth}) to each s.p. state $j$ coupled to continuum,
$W=\sum_j \gamma_j\,b^\dagger_j b_j .$ The same picture emerges
when the residual SM interaction is weak. Then s.p. motion masters
the dynamics, and the sum over daughter systems $\alpha$ in Eq.
(\ref{W1}) is dominated by a single term. This again leads to Eq.
(\ref{W2}) that is energy dependent, $W(E)=\sum_j
\gamma_j(E)\,b^\dagger_j b_j .$ The operator $W$ here can be
conveniently combined with the SM Hamiltonian just by introducing
complex s.p. energies for unstable orbitals in the mean field (a
simple instructive example was shown in \cite{VZCovello02}).
Another point to be mentioned here is related to the treatment of
the non-Hermitian part. In the full CSM diagonalization, virtual
transitions to continuum and real decays influence the internal
structure. This is particularly important at strong continuum
coupling when coherence with respect to decay leads to the
super-radiance phenomenon \cite{dicke54,sokolov89,volya_PRC67}.

The second situation leading to the SM picture is the limit of
weak continuum coupling when the matrix $W$ can be treated
perturbatively. In the lowest order, decays do not affect the
internal state and we can solve the Hermitian problem first to
obtain a many-body state $|\alpha\rangle$ with the width given by
the expectation value
\begin{equation}
\Gamma_\alpha=\langle \alpha |W| \alpha \rangle =\gamma_j
\Upsilon_j(\alpha).
\end{equation}
This expresses a many-body decay width as a product of the s.p.
width and the spectroscopic factor $\Upsilon_j(1)=\langle
1;N|b^\dagger_j b_j|1;N\rangle.$

\subsection{Two-nucleon emission\label{sec::2n}}

A two-body decay channel state is fixed by an $N-2$ nucleus in its
eigenstate $\alpha$ and a state of two nucleons in the continuum,
Eq. (\ref{twob}). These states are characterized by total energy
$E=E_\alpha+\epsilon_j+\epsilon_{j'}$ combined of the daughter
energy $E_\alpha$ and energies of emitted nucleons in s.p.
channels $j$ and $j'$, total angular momentum and isospin of the
emitted pair. Unless we are dealing with a bound two-particle
continuum state, the channel, besides total energy $E$, has
another continuous index describing the energy distribution
between the particles.

As earlier, the two-body transition amplitude is generated by the
matrix element $\langle 1|H|c;E\rangle $. Two different
contributions can be identified as ``direct'' and ``sequential'',
Fig. 2. The defined above one-body part of the total Hamiltonian
$H$ cannot contribute to the direct decay vertex $1 \rightarrow
\alpha + j + j'$. The two-body interaction responsible for this
direct transition is discussed below. Even without two-body
interactions, the ``dressed'' vertex is not zero since the $N-2$
daughter state is a part of the virtual cloud of the $N-1$ system
to which one-body transitions are allowed. Such a second order
perturbative sequential mechanism of decay is possible regardless
of whether one-body channel is open or closed. Similar approach
and classification of processes have been recently discussed in
the context of two-proton radioactivity \cite{rotureau05,brown03}
where on one hand the case is complicated by being a three-body
Coulomb problem \cite{grigorenko03}, while on the other hand
extremely weak decays do not affect the internal nuclear structure
allowing the use of the traditional real energy SM \cite{brown03}.
\begin{figure}[ht]
\begin{minipage}[1]{2.3 in}
\includegraphics[width=2.3 in]{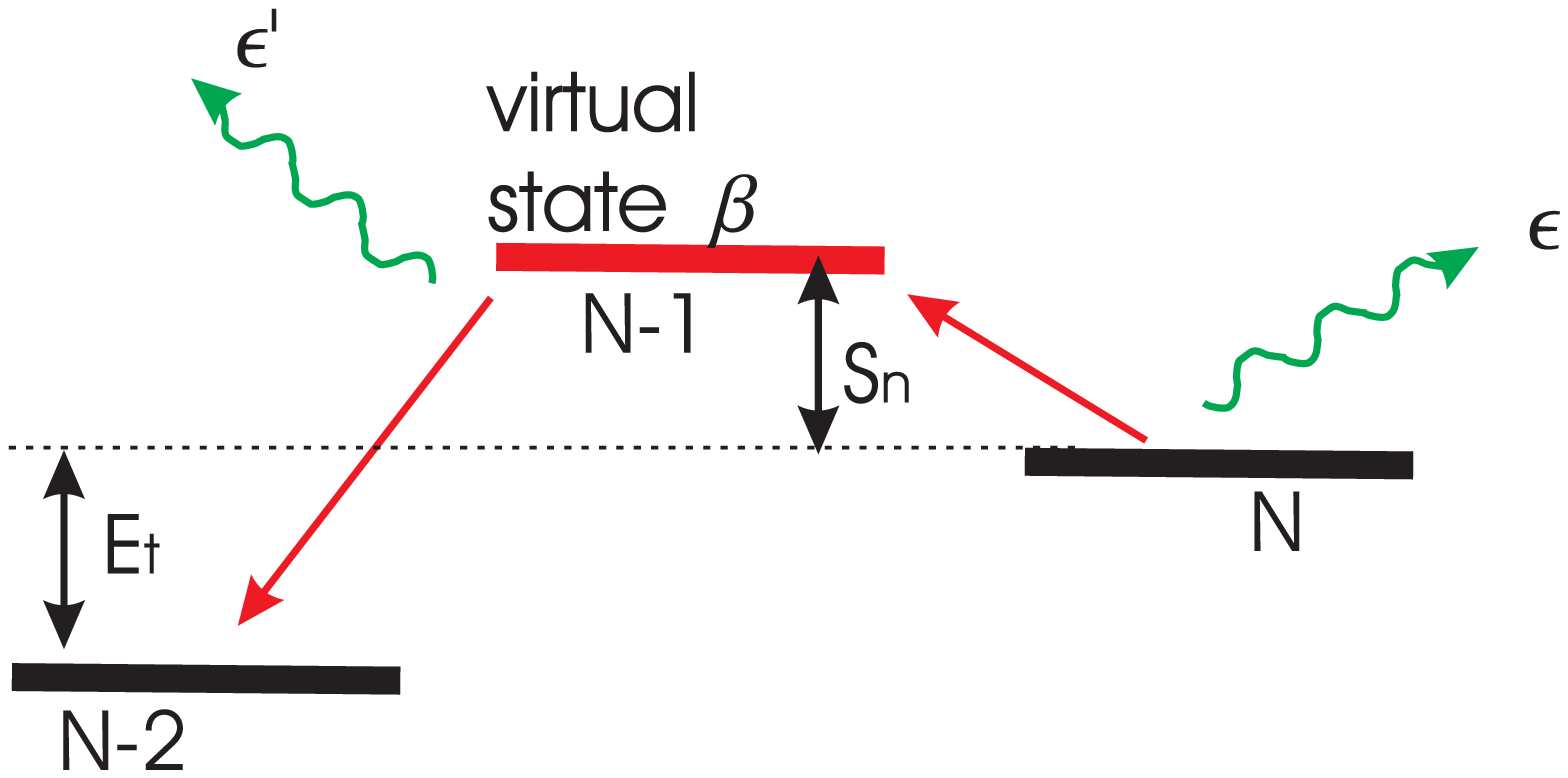}
\end{minipage}
\begin{minipage}{2 in}
\includegraphics[width=2 in]{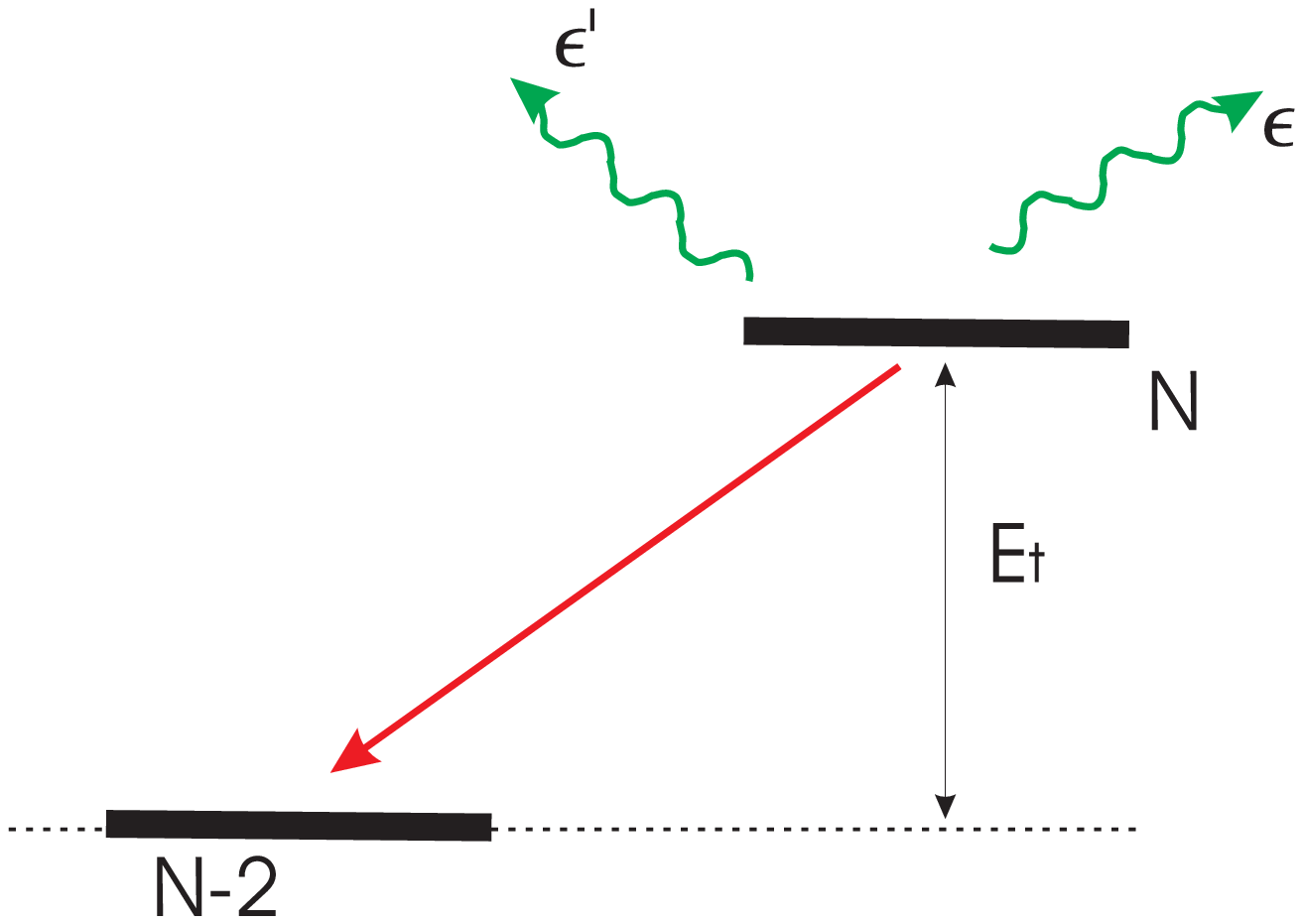}
\end{minipage}
\caption{(Color online) Diagrams for two-body decays, sequential
(left part) and direct (right part).  \label{decays}}
 \vskip -0.4 cm
\end{figure}

\subsubsection{Sequential decay}

The dressed s.p. vertex for two-nucleon emission $\langle
1|H|c;E\rangle $ can be calculated with the help of Eq. (\ref{2})
and solution (\ref{chi}),
\begin{equation}
A^c_1(E_c,E)=\sum_\beta \frac{A^{\{\beta
j\}}_1(E_\beta+\epsilon,E)\,A^{\{\alpha
j'\}}_\beta(E_\alpha+\epsilon',E_\beta)}{E_\beta-E_\alpha-\epsilon'}-\Bigl
( \{\epsilon, j\}\leftrightarrow \{\epsilon', j'\}\Bigr ).
\end{equation}
Assuming that internal and external s.p. states are orthogonal and
using one-body decay amplitude (\ref{onebcont}) for the second
order process that takes place via a virtual state $\beta$ being
suppressed by the energy barrier, we obtain
\begin{equation}
A^c_1(E)= \sum_\beta a_j(\epsilon) a_{j'}(\epsilon') \left
(\frac{\langle
1|b^\dagger_{j}|\beta \rangle\, \langle \beta|b^\dagger_{j'}|\alpha\rangle}
 {S_\beta+\epsilon}- \Bigl(\{\epsilon,
j\}\leftrightarrow \{\epsilon', j'\}\Bigr)\right )\,,
                                         \label{sequential}
\end{equation}
where we are considering only on-shell decay so that the initial
state with energy $E=E_\alpha+\epsilon+\epsilon'$ undergoes a
transition via the intermediate state with energy
$E_\beta+\epsilon$. This is reflected in the denominator
$S_\beta+\epsilon =E_\beta-E+\epsilon$, where $S_{\beta}$ stands
for separation energy from the intermediate system. Since $c$ here
contains a continuous index, the summation over channels is
expressed as
\begin{equation}
\langle 1|W(E)|2\rangle =
 \pi \sum_{\alpha,j,j'}\int\, d\epsilon\, d\epsilon'\,
\delta(E-E_\alpha-\epsilon-\epsilon')\, {A^c_1(E)} A^{c\ast}_2(E)
\,.                                     \label{sequential3}
\end{equation}
In order to avoid double counting due to fermion permutation we
included a factor of 1/2 while making domain of integration
symmetric. Eq. (\ref{sequential3}) may contain poles in open
one-body channels corresponding to real states of an $N-1$ system
through which the two-body decay process can proceed. Such
processes can be viewed as a second order correction to the
one-body decay via the daughter state $\beta$ as a resonant state.
Here we discuss only the off-shell contribution. An example of
sequential decay is shown in Fig.~\ref{decays}.

The calculation of the non-Hermitian term $W$ in the most general
case was carried out numerically in the examples shown in the
Sec.~\ref{sec::appl}. Here we illustrate the case of a weakly decaying
state $|\alpha_i\rangle$ when the decay width can be approximated
by the diagonal element, $\Gamma=\langle \alpha_i|W|\alpha_i
\rangle $. In this limit the problem is close to that of
sequential decay discussed in \cite{brown03,barker99}.

Taking into account the direct and exchange contributions to Eq.
(\ref{sequential}), assuming spherical symmetry and performing the
summation over magnetic quantum numbers, we obtain two
spectroscopic factors,
\begin{equation}
\Upsilon^{\rm d}=\delta_{j j'} \delta_{\beta \beta'}\,\frac{\left
|\langle \alpha_f||b_j||\beta\rangle \langle
\beta||b_j||\alpha_i\rangle \right|^2}{(2{\alpha_i}+1)(2\beta+1)},
\end{equation}
and
\begin{equation}
\Upsilon^{\rm x}=(-1)^{\beta +\beta'} \left \{
\begin{array}{ccc}
{\alpha_i} & j & \beta '\cr {\alpha_f} & j' & \beta \cr
\end{array}
\right \} \frac{ \langle \alpha_f||b_{j'}||\beta'\rangle^* \langle
\beta'||b_{j}||\alpha_i\rangle^* \langle
\alpha_f||b_j||\beta\rangle \langle \beta||b_{j'}||\alpha_i\rangle
} {2{\alpha_i}+1},
\end{equation}
where (and below) we use the Greek index of the many-body state as
a symbol of angular momentum. The reduced matrix elements are
defined as in \cite{BMI}. The total width in the channel
$|\alpha_i\rangle\rightarrow |\alpha_f \rangle$ is given by
\begin{equation}
\Gamma =\frac{1}{2 \pi} \int\, d\epsilon d\epsilon'\,
\delta(E_t-\epsilon-\epsilon')\,   \sum_{\beta \beta' j j'}
\gamma_j(\epsilon) \gamma_{j'}(\epsilon')\left (\frac{\Upsilon^{\rm
d}} {(S_\beta+\epsilon)^2} + \frac{\Upsilon^{\rm
x}}{(S_\beta+\epsilon')(S_\beta'+\epsilon)} \right)
\end{equation}

As an illustration, we discuss a case of the $0^+\rightarrow 0^+$
two-neutron decay. Here $\Upsilon^{\rm x}=\Upsilon^{\rm d}\equiv
\Upsilon$, and we can view a full width as a sum of partial widths
that depend on the final state $\alpha_f$ and intermediate state
$\beta$. The characteristic energies in the problem are the
one-neutron separation energy to the state $\beta$,
$S_\beta=E_\beta-E_{\alpha_i}$, and two-neutron decay energy,
$E_t=E_{\alpha_i}-E_{\alpha_f}$ (minus two-neutron separation
energy). We assume the situation with small available energy
$E_{t}=\epsilon+\epsilon'$, when the energy scaling of the  s.p.
decay widths, $\gamma_j=2\pi |a^j(\epsilon)|^2 \sim
\epsilon^{l+1/2}$, can be used. Introducing $q=E_t/S_\beta$ we
obtain
\begin{equation}
\Gamma_\beta(0\rightarrow 0)=\Upsilon \frac{\gamma_l(E_t)
\gamma_{l'}(E_t) E_t}{S_\beta^2} {\cal B}_{l l'}(q),
\end{equation}
which includes the phase space integral
\begin{equation}
{\cal B}_{l l'}(q)=\frac{(2+q)}{2\pi}\int_0^1\frac{x^{l+1/2}
(1-x)^{l+1/2}}{(1+qx)^2[1+q(1-x)]} dx\,.              \label{B}
\end{equation}
In the limit $S_n\gg E_t$ we obtain  ${\cal B}_{l l'}(q)=
{B}(l+3/2,l'+3/2)/\pi$, where $B(x,y)$ is a beta function. The
decay rate is suppressed by the energy denominator $1/S_\beta^2$.
The two-body decay width scales with energy as $E_t^2.$ This is
consistent with the phase space volume estimate, $\Gamma \sim
\delta^3(P_t-P)\delta(E_t-E) \prod_i (d^3 k_i/ \epsilon_i)$, where
$P$ is total momentum and the product $\Pi_i$ runs over the
fragment indices including the daughter nucleus. For the isotropic
case integrated over all angles, the width is proportional to
$\gamma \sim k\sim E_t^{1/2}$ in one-body decay; the same
assumptions lead to $\Gamma \sim E_t^2$ for the two-body decay
(three-body final phase space).

In the opposite limit, $S_n\ll E_t$ or $q \rightarrow \infty$, we
have ${\cal B}_{l l'}(q)= {B}(l-1/2,l'+1/2)/(2 \pi q^2)$. This
expression diverges for an $s$-wave. The exact integration in the
$s$-wave case gives
\begin{equation}
\Gamma_{\beta=1/2}(0\rightarrow 0)= \frac{\gamma_0^2(E_t)
\Upsilon_\beta E_t}{4 (2S_\beta+E_t)\, \sqrt{S_\beta(S_\beta+E_t)}}
. \label{swd}
\end{equation}
Note the divergence in the $s$-wave channel when separation energy
goes to zero. The $s$-wave state in the intermediate nucleus
$\beta$ is so broad that even being slightly higher in energy it
still poses no barrier for the sequential decay. The generic
behavior of the isotropic decay width as a function of energy can
be traced from Eq. (\ref{swd}), where $\Gamma\sim E_t^2$ if energy
is low, $E_t\ll S_\beta$,  but once energy is getting above
$S_\beta$ the behavior changes to $\Gamma\sim \sqrt{E_t}$. This
shows that the presence of the one-body resonance changes the
nature of sequential decay reflecting the one-body phase space
characteristics.

\subsubsection{Direct decay}

Direct two-body transitions are generated by the two-body part of
the Hamiltonian $H_{\cal PQ}$ that takes a nucleon pair coupled to
angular momentum $L$, $p_L=[b_j b_{j'}]_L$, from internal space
${\cal P}$ and transfers it to the two-body continuum ${\cal Q}$,
see the right part in Fig~\ref{decays}. The transition amplitude
has a generic form,
\begin{equation}
A^c_1(E)=a^{(j_1 j_2)}(\epsilon_1, \epsilon_2)\,\, \langle 1;N
|\left ( p^{(j j')}_L\right )^\dagger|\alpha;N-2\rangle\,,
\end{equation}
where a direct two-body transition amplitude, $a^{(j_1
j_2)}(\epsilon_1, \epsilon_2)$, is introduced (not to be confused
with s.p. amplitudes $a^j(\epsilon)$). This amplitude can be
calculated for a given two-body interaction. For example, assuming
for simplicity a coordinate form $V^{(2)}(r,r')$, where $r$ and
$r'$ are particle coordinates in the mean-field frame, the
amplitude can be expressed following the definition of Eq.
(\ref{ampa}) and normalized free particle states $F_j$ as
\begin{equation}
a^{(j_1 j_2)}(\epsilon_1 \epsilon_2)=\langle  j_1 j_2|V^{(2)}| j_1,
\epsilon_1; j_2,
\epsilon_2 \rangle = \frac{2 \mu}{\pi \sqrt{k_1
k_2}} \int_0^\infty dr dr' F_{j_1}(r) F_{j_2}(r') V^{(2)}(r,r')
u_{j_1}(r) u_{j_2}(r') \,.                       \label{2bspA}
\end{equation}

The low-energy behavior of the direct amplitude can be understood
without specification of the residual interaction by taking into
consideration the long wavelength behavior of the Bessel functions
associated with the regular solution, $F_l(r)\sim (kr)^{l+1}$ at
$k\rightarrow 0$. The decay rate can be estimated by integration
over continuous channel variables as in (\ref{sequential3}),
\begin{equation}
\Gamma\sim \int\, \left | a^{(j_1 j_2)}(\epsilon_1 \epsilon_2)
\right |^2\, \delta (E_t-\epsilon_1-\epsilon_2) \,d\epsilon_1\,
d\epsilon_2 \,   \sim E_t^{l_1+l_2+2}.
\end{equation}
The same answer as the one obtained for the sequential transition
reflects the nature of three-body final phase space. The direct
transition however is not suppressed by the one-body energy
barrier and is not related to decay amplitudes for one-body
decays.

We conclude the discussion of two-body decay processes with the
word of caution: the full amplitude in a given decay channel is
the sum of the direct and sequential contributions and the
observed width or cross section carries their interference.

\section{Applications\label{sec::appl}}

\subsection{How the method works}

Before turning to the discussion of the specific results we
outline the stages involved in the calculation. A special feature
of our approach is the exact treatment of threshold behavior. This
requires the knowledge of the topography of thresholds and
therefore has to rely on the preceding solution for the daughter
systems. In this way we come to problems of considering the
daughter chains in their entirety. Thus, except for the case of
the traditional SM, where the Hermitian Hamiltonian matrix is
diagonalized separately for each nucleus, in all calculations the
nuclides (in our examples below the isotopes) are coupled by the
decay chains.

The procedure starts from the closed core (specifically, $^{4}$He
or $^{16}$O) and continues toward heavier isotopes so that the
properties of all possible daughter nuclei are known prior to each
new calculation. The process of calculating resonant states is
iterative. We start with a given state obtained from the
conventional SM. For this state we review all possible decays,
including the two-body ones. By conducting scattering calculations
at relevant energy or via appropriate power-law dependence of Eq.
(\ref{plaw}) at low energies and including the spectroscopic
factors from structural rearrangement of spectator nucleons we
determine the contributions from each decay channel to the
non-Hermitian term $W$. The diagonalization of the full
Hamiltonian with the non-Hermitian part results in the next
iteration for energy and width of the state under consideration.
For each resonant state this process continues until convergence
is reached. The Breit-Wigner definition of resonances is used so
that scattering calculations are done at real energy.

By construction  of the model, with $\Delta(E)$, Eq. (11), assumed
to be included in the adjusted SM Hamiltonian with neglected
energy dependence, bound levels coincide with those in the
standard SM. For unbound states, this choice makes our internal
propagator equivalent to the $R$-matrix used in spectroscopic
analysis of experimental data. Therefore this design is most
suitable for the effective extraction of interaction parameters
from experiment.

The continuum coupling restructures internal states, and energies
of resonances above the decay threshold deviate from the SM
predictions. For narrow states and well separated resonances, the
resulting effect is small. The spherical shape of the semi-magic
system is stable; in addition, strong collective pairing reduces
the effect of decays onto internal structure in these particular
examples. Whence, for s.p. decays, the case (a) in
Table~\ref{hetab}, the use of the spectroscopic factor
approximation discussed in Sec.~\ref{sec::ob} is rather good. Here
we are far from strong coupling to continuum
\cite{pentaquark,rotter01,JOB5,sokolov89} that may cause an
internal phase transition with formation of broad (super-radiant)
and very narrow (trapped) states. In nuclear physics this
phenomenon separates compound and direct reactions
\cite{teichmann50,teichmann52,sokolov89}. A trace of this effect
is seen in Table~\ref{hetab}, where decaying states have their
resonant energy shifted from SM prediction, the lowest states of
given quantum numbers are pushed into continuum
\cite{volya_PRC67}.

The cross section calculations follow directly the formalism
outlined in Sec. \ref{cross}. The large scale repetitive matrix
inversion and instability in the vicinity of the poles presents a
significant technical challenge. In this work we employed a new
numerical method that expands the unperturbed propagator (\ref{G}) in
the time domain using Chebyshev polynomials. The entire procedure
is similar to the Lanczos technique and involves only matrix
vector multiplication - a fast operation with sparse matrices. The
resulting $R$ matrix is then used in the Woodbury equation
(\ref{woodbury}) leading to a full propagator and scattering cross
section.

\subsection{Helium isotopes}

For the chain of helium isotopes from $^4$He to $^{10}$He, the
results are summarized in Table. \ref{hetab}. The internal valence
${\cal P}$-space contains two s.p. levels, $p_{3/2}$ and
$p_{1/2}$, the $\alpha$-particle core is kept inert. The effective
interaction within this model space and s.p. energies are borrowed
from \cite{cohen65,stevenson88}. Without additional terms, this
would be merely a conventional SM leading to the bound states that
are listed as E(SM) in Table~\ref{hetab}.

\begin{figure}
\begin{center}
\includegraphics[width=11 cm]{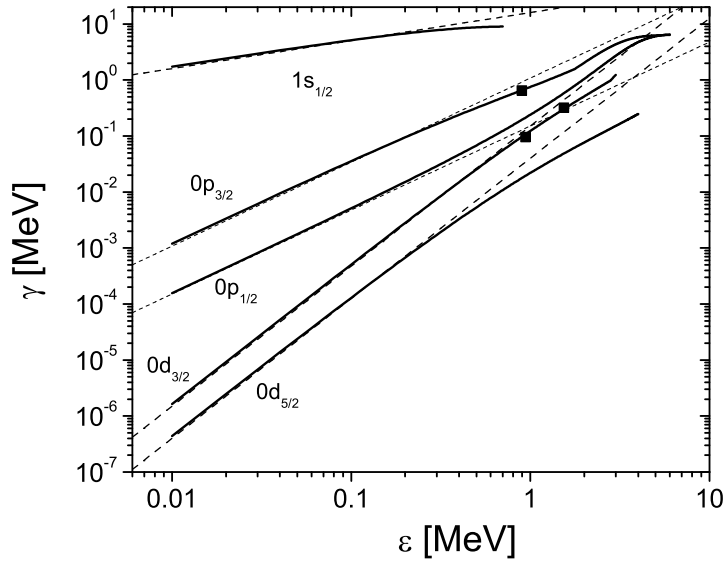}
\end{center}
\caption{Single-particle decay width as a function of resonance
energy for the $p$ and $sd$ shell models. 
The solid lines represent results of the
reaction calculations based on the
Woods-Saxon potential appropriate for the corresponding case 
(helium or oxygen)
discussed in text. One experimental point on the $p$-state 
corresponds to the ground state  $3/2^-$ resonance in $^5$He
($\epsilon=0.895$ MeV and $\gamma=648$ keV). 
Two points for the $d$-state correspond to
experimentally known $3/2^+$ resonances in $^{17}$O
($\epsilon=0.941$ MeV and $\gamma=98\pm 5$ keV) and $^{19}$O
($\epsilon=1.540$ MeV and $\gamma=320\pm 25$ keV
\cite{skorodumov05}). The near-threshold power law fits of Eq.
(\ref{plaw}) are shown with dashed lines. \label{weo16wss}}
\end{figure}

The one-body part of the coupling Hamiltonian $H_{\cal PQ}$ is
defined using the Woods-Saxon potential. However, at low energies 
the s.p.
decay widths can be well approximated by near-threshold dependence
of Eq. (41) as $\gamma_{3/2}(\epsilon)= 1.08
\,\epsilon^{3/2}\,$MeV and $\gamma_{1/2}(\epsilon)=0.151\,
\epsilon^{3/2}\,$MeV for $p_{3/2}$ and $p_{1/2}$ states,
respectively, see Fig. \ref{weo16wss}. The results
of calculation limited to s.p. decays are shown as case (a) in
Table~\ref{hetab}. The ``Borromean" $^6$He nucleus requires
two-body decays. The sequential two-body decay as a second order
process built on the one-body amplitudes involves no additional
parameters. This process is included in the results shown as case
(b) in Table~\ref{hetab}.

The judgment about the direct two-body decays, their importance,
and whether two-body terms have to be included in $H_{\cal PQ}$
can be made with comparison of theory versus experimental data.
Considering the simplest two-body case of $^6$He we observe that
the included sequential decay provides a reasonable description
for the lowest $2^+_{1}$ resonance in $^6$He. However, the need to
include direct two-body decay comes from consideration of a broad
asymmetric resonance structure at excitation energy $\sim 3-15 $
MeV \cite{janecke96,nakayama00} ascribed to neutron skin
oscillations. Although an experimental controversy exists in
relation to this structure, our results below, as well as earlier
studies, see \cite{janecke96}, suggest a direct two-body breakup
process. Based on the experimental exclusion of $2^+_2$
possibility by \cite{nakayama00}, we tentatively identify this
resonance as a spin zero pair-vibration excitation generated by a
coherent transition of the $L=0$ neutron pair between the two s.p.
orbitals and introduce direct two-body decay with the pair
emission in the $L=0$ channel. At this level of experimental
precision and with the sequential decays being already included,
there is no compelling evidence for direct decays in other angular
momentum channels.

\begin{figure}
\begin{center}
\includegraphics[width=5 in]{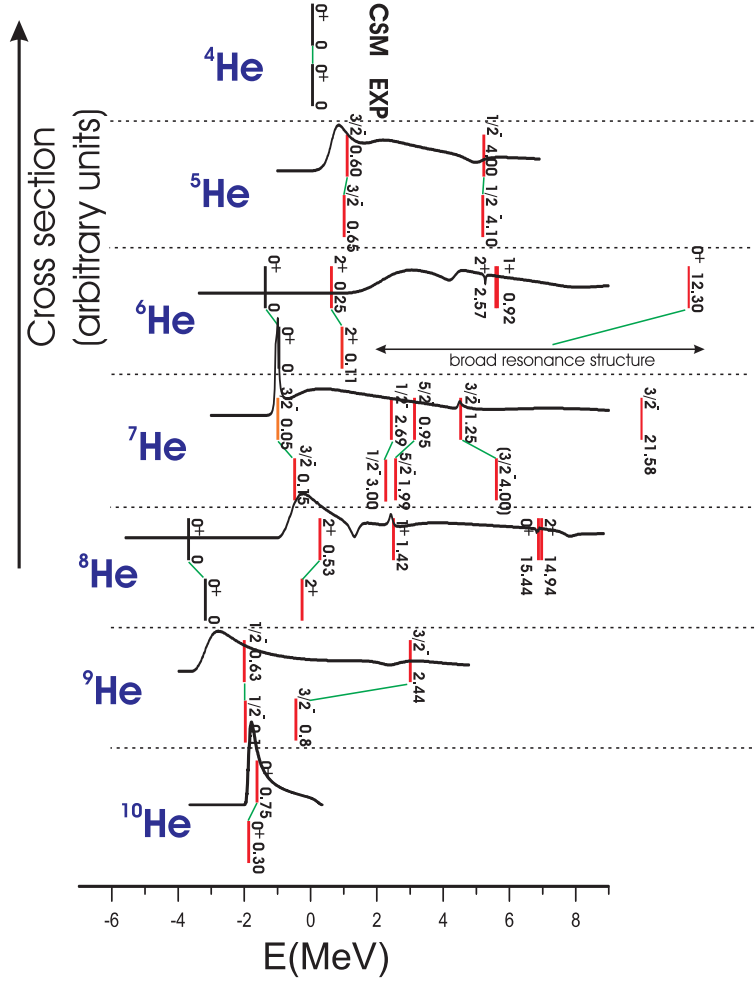}
\vspace{-0.2 cm}
\end{center}
\caption{(Color online) CSM results for He isotopes. The states in
the chain of isotopes starting from $^4$He (top) to $^{10}$He
(bottom) are shown as a function of the energy relative to $^4$He.
The horizontal dotted lines separate each isotope and for each
case states from CSM are shown above experimentally observed
states. The decay width (in MeV) is shown for each state along
with spin and parity. The solid lines above CSM states shows the
elastic neutron scattering cross section from the spin polarized
state of $N-1$ isotope, see text for details. \label{he2}}
\vspace{-0.2 cm}
\end{figure}

Without introducing a specific isotropic form of the two-body
interaction for $H_{\cal PQ}$ we adopt a low-energy ``pairing"
approximation assuming that all $L=0$ neutron pairs couple to
continuum with the same amplitude $a^{(L=0)}(\epsilon_1,
\epsilon_2)= (\epsilon_1+\epsilon_2)/{3\sqrt{2\pi}}$. The
numerical constant here sets the strength of the residual two-body
interaction. This is the only parameter of the model, and with
limited experimental knowledge it was selected to roughly
reproduce the broad resonant structure in $^6$He. The results of
the full calculation are shown in Fig.~\ref{he2} and listed as
(CSM) in Table~\ref{hetab}. Fig.~\ref{he2} also shows elastic
neutron scattering cross section from the ground state of the
$N-1$ nucleus in $M=0$ (even) or $M=1/2$ (odd mass) magnetic
moment state. The cross section is computed with the same CSM
parameters but within the case (a) discussed above.

\begin{table}
{
\begin{tabular}{|c|c||c|c|c|c|c||c|c|c|c|}
\hline
A&   J &     E(SM) &    E(a)&  E (b)&  E(CSM)&    E(EX)&  $\Gamma$(a) &  $\Gamma$(b)& $\Gamma$(CSM)&    $\Gamma$(EX)\\
\hline
4&  0&  0&  0&  0&  0&  0&  0&  0&  0&  0\\
5&  3/2&    0.992&  0.992&  0.992&  0.992&  0.895& 0.6&    0.6&    0.6&    0.648\\
5&  1/2&    4.932&  4.932&  4.932&  4.932&  4.895& 4&  4&  4&  4.1\\
6&  0&  -1.379& -1.379& -1.379& -1.379& -0.973&    0&  0&  0&  0\\
6&  2&  0.515&  0.515&  0.529&  0.529&  0.825& 0&  0.248&  0.248&  0.113\\
6&  2&  4.745&  5.25&   5.25&   5.25&   &   2.566&  2.566&  2.566&  \\
6&  1&  5.889&  5.32&   5.32&   5.32&   &   0.922&  0.922&  0.922&  \\
6&  0&  11.088& 10.911& 10.911& 10.803&   & 5.532&  5.532&
12.303&
\text{broad}\\
7&  3/2&    -1.016& -1.016& -1.016& -1.016& -0.528&    0.046&  0.046&  0.046&  0.15\\
7&  1/2&    2.24&   2.239&  2.253&  2.253&2.172  &   2.357&  2.689&  2.69& 3  \\
7&  5/2&    2.85&   2.888&  2.911&  2.911&  2.393& 0.727&  0.944&  0.944&  1.99\\
7&  3/2&    4.495&  4.379&  4.222&  4.22&   (5.273)& 0.541&  1.113&  1.246& (4)\\
7&  3/2&    10.223& 8.857&  9.521&  9.544&  &   7.818&  16.379& 21.578& \\
8&  0&  -3.591& -3.591& -3.591& -3.591& -3.108&    0&  0&  0&  0\\
8&  2&  0.19&   0.196&  0.191&  0.19&   -0.308&    0.231&  0.506&  0.53&   \\
8&  1&  2.427&  2.304&  2.331&  2.321&  &   1.026&  1.455&  1.418&  \\
8&  0&  6.376&  6.003&  6.527&  6.489&  &   5.286&  3.456&  15.449& \\
8&  2&  6.882&  6.839&  6.538&  6.572&  &   2.283&  13.86&  14.94&  \\
9&  1/2&    -1.992& -1.992& -1.992& -1.992& -1.958&    0.634&  0.634&  0.634& 0.1 \\
9&  3/2&    2.805&  2.801&  2.802&  2.797& -0.6  &   1.557&  2.425&  2.443& 0.8 \\
10& 0&  -1.649& -1.649& -1.649& -1.649& &   0.073&  0.504&  0.746& 0.3 \\
\hline
\end{tabular}
} \caption{\label{hetab}Comparison of conventional SM and CSM with
data for He isotopes (all numbers in MeV; energies are measured
from the ground state of $^4$He). The first two columns indicate
the mass number and spin of the state. The next five columns
compare energies as follows: $E$(SM) - traditional shell model;
$E$(a) - CSM with only one-body decays included; $E$(b) - CSM with
one-body decay and its second order contribution to the two-body
process; $E$(CSM) - full CSM including the direct two-body decay
mode; $E$(EX) - experimental data (some of them have large
uncertainties and depend on the way of analysis). Last columns
compare decay widths from CSM calculations with data
\cite{nndc,rogachev04,korsheninnikov99}; the SM calculation gives
only discrete energies.} \vspace{-0.2 cm}
\end{table}

Below we comment on some of the features of the results.

(i) The ground states of $^{4,6,8}$He are nucleon-stable in
agreement with experiment.

(ii) The results reveal information about structure and dominant
decay modes. For example, for the $^7$He isotope, that has not
been used for an adjustment of interactions, they agree with
recent experiments \cite{rogachev04,korsheninnikov99}. Our results
support the ``unusual structure" of the $5/2^-$ state identified
by \cite{korsheninnikov99}. Due to its relatively high spin, this
state, unlike the neighboring $1/2^-$ state, decays mainly to the
$2^+$ excited state in $^6$He. Concerning the deviations from the
data in the width of this state and $3/2^-$ state, we may comment
that experimentally \cite{korsheninnikov99}, although there are
still serious uncertainties, it is most likely that the
contribution from the triple decay sequence, sequential two-body
followed by one-body, is important for the lifetime of the $3/2^-$
state. The processes of the third order one-body decay or second
order combination of direct two-body followed by one-body, are not
included in this calculation but in light of the experimental data
they should be considered in the future.

(iii) Our results and, most importantly, the theoretical
interpretation of the role that one-body and two-body interaction
terms play in the dynamics of decay agree with earlier findings in
\cite{michel03} where a different method involving a full
diagonalization without separation of continuum into channels was
used.

(iv) For the heaviest isotopes and, in particular, for the
discussion of the presently interesting case of $^{9}$He, the
extension to the $sd$ shell is necessary.

(vii) The scattering cross section is non-zero only above the
thresholds (the ground state of the $A-1$ nucleus). Cross section
curves are generally not symmetric and have neither Gaussian nor
Lorentzian shape. For low-lying states with the width big enough
to reach threshold, the distortion is particularly noticeable.

(viii) For broad resonances, the identification of resonances (or
poles of the scattering matrix) with a peak in the cross section
is ambiguous. Interference between different resonances, including
rather remote once, is significant in this case.

\subsection{Oxygen isotopes}

The internal space here is represented by the $s_{1/2}$, $d_{3/2}$
and $d_{5/2}$ s.p. orbitals composing the usual $sd$-shell model.
The standard SM interactions (USD \cite{USD} or its slightly
modified version for heavier isotopes HBUSD \cite{HBUSD}) were
used in these calculations.

Since we use the Woods-Saxon potential for reaction calculations,
it is important to make sure that the potential create proper
resonant states. For few low-lying s.p. resonances with large
spectroscopic factors, we adjust the depth of the potential so
that the correct s.p. resonance energy is indeed ensured. The
Woods-Saxon potential parameterization with the mass dependent
depth \cite{skorodumov05,goldberg04} was demonstrated to reproduce
the s.p. resonances and bound states with good precision for
nuclei around mass $A=16$. For the majority of high-lying states,
the potential is not readjusted, here the decay amplitude is
computed directly from Eq. (\ref{spA}). The resonance phenomenon
in this case is due to the many-body effect where the complexity
of the many-body states makes the overlap in Eq. (\ref{onebcont})
small.

In Fig. \ref{weo16wss} the widths of the resonant states
$s_{1/2}$, $d_{3/2}$ and $d_{5/2}$ are shown as a function of their
energy being found with the aid of the Woods-Saxon potential with
variable depth. The curves are limited to the near-threshold
region approximately determined as $k R< l$ (and $kR\ll 1$ for
$l=0$) where $R$ is the nuclear radius. For oxygen, this limits
the $d$-wave at about 3 MeV. The curves are close to straight
lines displaying the appropriate power law scaling (\ref{plaw}) of
the decay width as a function of energy. The lines can be fit by
equations (with $\epsilon$ in units of MeV) $\gamma(s_{1/2})=16
\epsilon^{1/2}$, $\gamma(d_{3/2})=0.15 \epsilon^{5/2}$, and
$\gamma(d_{5/2})=0.04 \epsilon^{5/2}$. These fits are shown with
dashed lines.

\begin{figure*}
\begin{center}
\includegraphics[width=7 in]{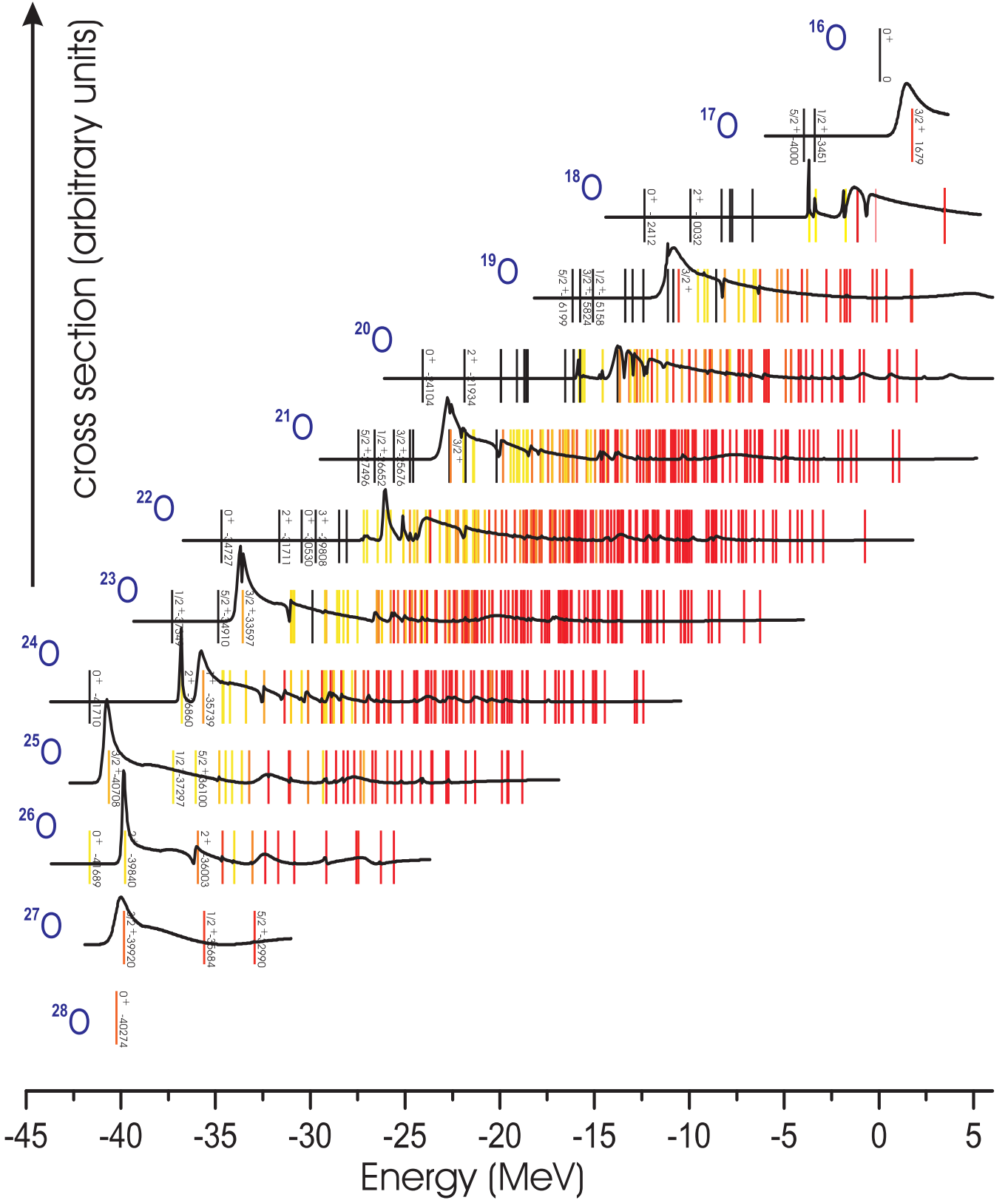}
\vspace{-0.2 cm}
\end{center}
\caption{(Color online) CSM calculation for oxygen isotopes with
the HBUSD interaction. The states (vertical bars) are plotted as a
function of absolute energy computed relative to ground state
$^{16}$O. States shown with solid black lines are stable in our
model; they are either below decay thresholds or with decays
forbidden due to the angular momentum restrictions in the selected
valence space. States from yellow (color online)/lighter shade of
gray (long lifetime) to red/darker gray (short lifetime) are
resonance states. For some of the low-lying states the energy (in
units of keV), spin and parity are given. Along with the resonant
level structure above thresholds, the elastic $d$-wave neutron
scattering cross section, off a ground state of a daughter nucleus
with $M=0$ (even mass) or $M=1/2$ (odd mass) magnetic quantum
number, is plotted as a function of energy. The cross section is
obtained as a part of the same CSM calculation. \label{os2} }
\vspace{-2 cm}
\end{figure*}

\begin{figure}
\begin{center}
\includegraphics[width=11 cm]{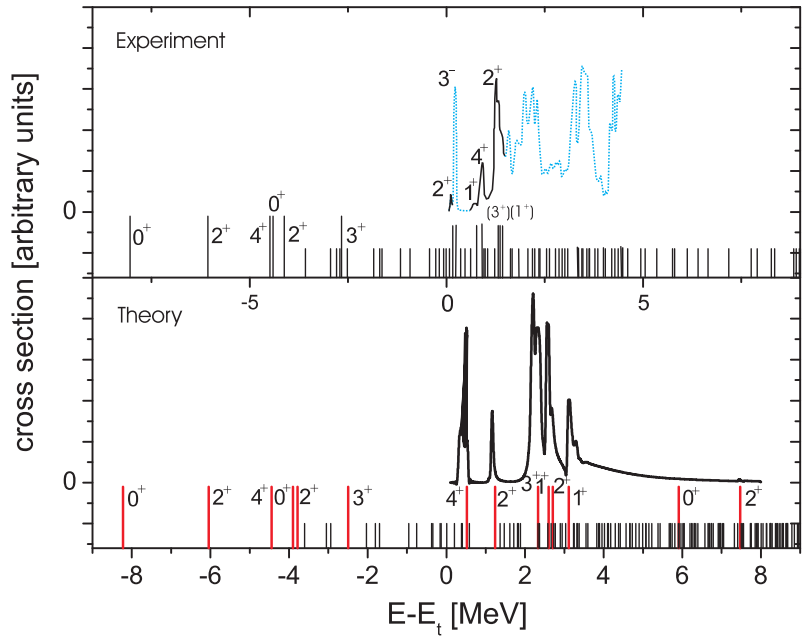}
\end{center}
\caption{(Color online) The lower panel is the result of the CSM
calculation of the level structure and neutron scattering cross
section from the ground state of $^{17}$O. Marked levels
corresponding to the $sd$-shell model states are included in CSM
calculation. All other levels result from the expanded $sd-pf$
shell calculation (see text); the configurations of this type are
not coupled to continuum and corresponding (negative parity)
states do not appear in the cross section plot. The upper panel
corresponds to the empirical level scheme in $^{18}$O. The neutron
scattering cross section on the upper panel is indirectly inferred
from the $^{14}$C$(\alpha n)^{17}$O reaction \cite{bair66} and
other compilations \cite{neutron,neutron2,tunl}. The parts of the
cross section shown with solid line are expected to be dominated
by the positive parity $l=0$ or $l=2$ partial waves and thus can
be distantly juxtaposed to the curve on the lower plot.
\label{o18cross}}
\end{figure}

In Fig.~\ref{os2} we show an overview of the full CSM calculation
for oxygen isotopes within the $sd$ shell. Some of the first
results were reported earlier \cite{volya_PRL,volya_PRC67}.

\begin{table}
\begin{tabular}{|c|c|c||c|c|c||c|c|c|c|}
\hline
    A&j & mode & EXP& Q & $\Gamma$ &theory & E &Q & $\Gamma$\\
\hline
17 & $3/2^+$ & $\gamma\, n $& 5.085 & 0.941 & 96 & WS & 4.5 & 1.0 & 122 \\
18 & $2^+$ & $\gamma\, \alpha\, n$ & 8.213 & 0.169 & $1 \pm 0.8$& USD & 9.465 & 1.242 & 200 \\
18 & $1^+$ & $ \alpha\, n$ & 8.817 & 0.773 & $70 \pm 12$& USD & 10.823    & 2.600 & 85 \\
18 & $4^+$ & $\gamma\, \alpha\, n$ & 8.955 & 0.911 & $43\pm 3$& USD & 8.750 & 0.526 & 28 \\
19 & $5/2^+$ & $n$ & 5.148 & 1.191 & $3.4\pm 1$ & USD & 5.011 & 1.121 & 5.1 \\
19 & $9/2^+$ & $ $ & 5.384 & 1.427 & $\sim 0 $ & USD & 5.175 & 1.282 & 0 \\
19 & $3/2^+$ & $n$ & 5.54 & 1.58 & $320$ & USD & 5.529 & 1.636 & 290 \\
19 & $7/2^+$ & $n$ & 6.466 & 2.509 & small & USD & 6.880 &
0.808\footnote{the $Q$-value is measured for the excited state in
daughter system.} & 63 \\
24 & $2^+$ & $n$ & ? & ? & ? & HBUSD & 4.850 & 0.489 & 18 \\
26 & $0+$ & $2n$ & 0 & ? & ? & HBUSD & 0 & 0.021 & 0.02 \\
28 & $0+$ & $2n$ & 0 & ? & ? & HBUSD & 0 & 0.345 & 14 \\
\hline
\end{tabular}
\caption{Lowest resonance states in the chain of oxygen isotopes.
The experimental data on the left, (EXP) - energy of the state
(MeV), Q - energy above threshold (MeV), $\Gamma$ - width (keV),
are compared to the theoretical results on the right. The decay
mode in the second column indicates the decay branches assumed by
experimentalists~\cite{tunl,nndc}.  \label{otab}}
\end{table}

Our calculations are in a good agreement with available
experimental data for the oxygen isotopes \cite{tunl,nndc}. One
has to emphasize again that, with the assumption of the self
energy term $\Delta$ being a part of the conventional SM
Hamiltonian, the bound states are exactly the same as obtained in
the usual SM calculation. The novelty appears when the states
above decay thresholds are considered. The properties of a few
experimentally identified resonances in oxygen isotopes are
compared to the CSM predictions in Table \ref{otab}.

There are two cases in $^{17}$O and $^{19}$O corresponding to the
neutron emission from the $d_{3/2}$ orbital. These rather pure
s.p. decays involve little many-body physics and are well
described with the Woods-Saxon potential model, Fig.
\ref{weo16wss}. Here the SM may be used to predict the neutron
decay energy, however there is an obvious improvement if the
experimental $Q$-value is used in the s.p. reaction calculations.
The power law for the energy dependence of the decay width can
enhance the predictive power of the description.

The two-body case of $^{18}$O is more complicated, however there
are only 14 states in the $sd$ shell model. The comparison of the
level structure and neutron scattering cross section with data is
shown in Fig. \ref{o18cross}. The SM below neutron decay threshold
(zero on the plot) is in an impressive agreement with the data.
Here, for all $sd$ SM states shown with longer ticks (in red,
color online), the experimental counterpart can be easily
identified in the observed spectrum. Practically all remaining
observed negative parity states below neutron threshold can be
identified with particle-hole excitations using an extended
$p-sd-pf$ SM. In theoretical calculation, these states are shown
with small lines (the WBP cross-shell interaction from Ref.
\cite{oxbash} was used).

Due to the resonant nature of states, the experimental picture
above neutron decay threshold becomes more ambiguous. The lowest
near-threshold state $2^+$ may correspond to one of the
higher-lying $2^+$ states in our model, see Fig. \ref{o18cross}
and Table \ref{otab}. Due to the significant difference in energy
above threshold between this state and its possible theoretical
counterpart, comparison of the decay widths is inappropriate.
Following this state, there are rather narrow 4$^+$ and $1^+$
states (the experimental assignment of $1^+$ for the latter state
is still uncertain), see Table \ref{otab} and Fig. \ref{o18cross}.
Although in theoretical calculation these states appear in a
different order their observed and calculated widths appear to
agree. Beyond this point there are several broad 100-200 keV
unidentified resonances observed in experiment. They may be
juxtaposed to 150-400 keV wide $2^+,\,3^+,$ and $1^+$ states
appearing in the CSM calculation. To support this argument, we
show on the lower panel of Fig. \ref{o18cross} the CSM neutron
scattering cross section computed using Eq. (\ref{scatt}). Since
this calculation takes into account only $l=0$ and $l=2$ partial
waves, the peaks from negative parity states do not appear. On the
upper plot we roughly reproduce the experimental cross section
obtained from indirect reactions, Refs.
\cite{bair66,neutron,neutron2,tunl}. The portion of the curve
corresponding to positive parity resonances that are expected to
be present in the theoretical calculation is shown by the solid
black curve. There is a reasonable similarity in the resonant
structure corresponding to the set of wide states $2^+,\,3^+,$ and
$1^+$. In the experiment, the spin-parity assignments for the
states shown in Fig. \ref{o18cross} in brackets have not yet been
confirmed.

The cross section picture indicates that our interpretation is
plausible, while differences, by the factor of 2 to 4, in the
resonance parameters may be due to different definitions of the
width used in the data analysis. The broadness of these resonances
causing difficulties in experimental analysis and interpretation
of peaks is here a trivial consequence of the simple two-particle
structure of $^{18}$O that implies high spectroscopic factors for
s.p. decay. The situation changes in the next isotope $^{19}$O
where there is only one broad s.p. $3/2^+$ state that stands out
in the calculations of the cross section. The majority of other
states are narrow due to the many-body complexity. They would not
be visible on the plot of the cross section; on the other hand,
this allows for the direct comparison of widths and energies shown
in Table \ref{otab}. The level scheme inferred from experiments
with $^{19}$O fits well the overall picture of oxygen isotopes as
shown in the insert of Fig. \ref{weo16wss}.

Only little is known about the heavier oxygen isotopes although
the situation is going to change in the near future with the new
radioactive beam experiments. In Table \ref{otab} we quote some,
in our view the most interesting, predictions of the model. The
cases of $^{26}$O and $^{28}$O are just beyond the drip line. The
ground states of these nuclei are unstable with respect to the
neutron pair emission. Sec.~\ref{sec::2n}, while being centered
around interactions, discussed how to tackle such cases.
Unfortunately, the uncertainty in the effective interactions
complicates the job of predicting. We have no firm knowledge about
the interaction coupling internal states to the continuum and,
unlike the helium case, in the oxygen data we could not find a
case to determine this coupling phenomenologically. Thus we only
consider here the case of sequential decay driven by what we
believe to be a well defined one-body potential.

The second problem is that the traditional SM adjusted to
experimental data near the stability line cannot be extrapolated
with full certainty to the vicinity of drip lines. The role of the
self-energy term $\Delta(E)$ is another question. Indeed, the well
established USD interaction predicts $^{26}$O to be bound which is
known from experiments not to be the case. In our calculations we
had to resort to the HBUSD interaction specifically adjusted to
heavier isotopes. The $Q$-value coming from this interaction in
the case of $^{26}$O is only 21 keV while the typical SM
uncertainty in level energies is about $200$ keV. This uncertainty
entering the power law scaling of decay width versus energy makes
the lifetime predictions unreliable.

Thus, while collecting our calculations in Table \ref{otab} we see
our strongest prediction through Eq. (\ref{swd}). In both cases of
$^{24}$O and $^{26}$O, the lowest $3/2^+, 1/2^+$ and $5/2^+$
states in the adjacent isotopes $^{23}$O and $^{25}$O are the main
candidates for the intermediate states involved in the sequential
two-body decay. This is because of the low energy barrier in
combination with large spectroscopic factors of these mainly s.p.
states. At the same time, in the case of $^{24}$O we have
separation energies $S_{3/2}=0.98,\, S_{1/2}=4.39$ and
$S_{5/2}=5.54$ MeV; the corresponding spectroscopic factors are
$\Upsilon_{3/2}=0.28,\, \Upsilon_{1/2}=0.019$ and
$\Upsilon_{5/2}=2.2\times 10^{-4}$. In addition to spectroscopic
information, the kinematics of the phase-space volume is another
essential factor. A simple estimate with the aid of Eq.
(\ref{swd}) shows that the transition through the $s$-state would
dominate throughout the entire region below s.p. threshold. Due to
level energetics, two-body decay remains significant even in the
presence of the open one-body channel. However, the simple power
law scaling used to obtain the phase space integrals Eq. (\ref{B})
may no longer be valid high above threshold. If the ground state
in $^{24}$O were at $E_t=1$ MeV above two-neutron decay threshold,
while still right at the opening of one-neutron decay channel into
the $3/2^+$ state of $^{23}$O, $S_{3/2}=0$, the decay width coming
from sequential decay would be about 30 keV, as opposed to 20 eV
in Table \ref{otab} quoted for $E_t=0.021$ MeV.

\section{Conclusion}

The first goal of this work is to present a systematic and
detailed discussion of the continuum shell model as a step in the
direction of unifying the nuclear structure with nuclear
reactions. We amplify and extend the ideas and methods started in
our earlier works \cite{volya_PRL, volya_PRC67}. In this
presentation we clarify the CSM formalism and show its relation to
the standard SM. One of the important points of this work was
physical interpretation of various results. On the shell model
side, we highlighted the meaning of solutions of the
energy-dependent non-Hermitian effective Hamiltonian and
identified the procedures to be taken in relation to different
definitions of resonant states. On the reaction side, we show how
the scattering matrix, cross sections and related quantities can
be calculated; the unitarity properties of the scattering matrix
built in the model are emphasized.

The general discussion of s.p. decays centers around the potential
s.p. problem. This textbook problem is not only a central part of
calculations but it also provides an important parallel to the
full CSM description. Here the Gamow states, decay amplitudes and
scattering matrix can be calculated through numerical solutions of
the Schr\"odinger equation in coordinate space. This allows one to
establish a transparent relation and interpretation of the same
quantities in the full CSM.

The consideration of two-body decays is one of the significant
advances of the present formulation. Considering the one- and
two-body terms in the part of the Hamiltonian that links the
internal shell model and external reaction space we get keys to
the sequential and direct decays, respectively. In the case of
sequential decay, a second order one-body process, we discuss the
transition through the resonance tail, namely the role of broad
one-body resonances located in the intermediate nucleus above
threshold. The direct, or correlated, decay processes are strongly
related to the problem of pairing and other coherent effects in
the continuum. This problem, important also for physics of neutron
stars, is still far from being solved.

The last section of this work shows practical applications of the
model. The self-consistency in energy, proper open channels and
realistic reaction calculations, parent-daughter structure
relations through the decay chain, $-$ are discussed as the
essential elements of the model. The diagonalization of the full
Hamiltonian with both Hermitian and non-Hermitian parts is
emphasized as an important component for treating properly the
mutual influence of structure and reactions. The discussion of the
extreme effects of this nature, such as super-radiance, is outside
the scope of this paper, see \cite{sokolov89,pentaquark,JOB5}.

Another goal of this work is to report the advance in practical
applications of the CSM and in particular to demonstrate new cross
section calculations performed in the same unified framework and
presented along with the bound states and resonance
parameterizations. Comparison with experimental data indicates a
satisfactory agreement for both helium and oxygen isotopes. The
possible role of sequential decay in heavy oxygen isotopes is
discussed and predictions for the decay width are given. Coming
experimental data will be instrumental for further development of
theory.

We did not discuss here the computational problems which are of a
higher level of difficulty compared to the normal SM. In fact, the
novel methods for constructing Green's functions for large-scale
calculations using the Chebyshev polynomial expansion and Woodbury
equation stand behind the presented results. However, the
technical and numerical details are outside the scope of this
article; they will be presented elsewhere. From the conceptual
viewpoint, the basic question of effective interactions remains
unsolved. We used here a semi-empirical method of combining the SM
experience with finding the missing cross-space matrix elements
from the solution of the scattering problem and numerical fits
based on general requirements of quantum-mechanical threshold
behavior. The inclusion of giant resonances, more complicated
decay modes and cluster channels is also on the agenda for future
work.

\section*{Acknowledgments}

The authors acknowledge support from the U. S. Department of
Energy, grant DE-FG02-92ER40750; Florida State University FYAP
award for 2004, and National Science Foundation, grants
PHY-0070911 and PHY-0244453. Help and useful discussions with B.A. Brown and
G. Rogachev are highly appreciated.


\begin{thebibliography}{55}
\expandafter\ifx\csname natexlab\endcsname\relax\def\natexlab#1{#1}\fi
\expandafter\ifx\csname bibnamefont\endcsname\relax
  \def\bibnamefont#1{#1}\fi
\expandafter\ifx\csname bibfnamefont\endcsname\relax
  \def\bibfnamefont#1{#1}\fi
\expandafter\ifx\csname citenamefont\endcsname\relax
  \def\citenamefont#1{#1}\fi
\expandafter\ifx\csname url\endcsname\relax
  \def\url#1{\texttt{#1}}\fi
\expandafter\ifx\csname urlprefix\endcsname\relax\def\urlprefix{URL }\fi
\providecommand{\bibinfo}[2]{#2}
\providecommand{\eprint}[2][]{\url{#2}}

\bibitem[{\citenamefont{Zelevinsky}(2005)}]{WNMP04}
\bibinfo{editor}{\bibfnamefont{V.}~\bibnamefont{Zelevinsky}}, ed.,
  \emph{\bibinfo{title}{Nuclei and Mesoscopic Physics}}, vol.
  \bibinfo{volume}{777} of \emph{\bibinfo{series}{Conference Proceedings}}
  (\bibinfo{publisher}{AIP}, \bibinfo{year}{2005}).

\bibitem[{\citenamefont{Brown}(2001)}]{brown01}
\bibinfo{author}{\bibfnamefont{B.A.}~\bibnamefont{Brown}},
  \bibinfo{journal}{Prog. Part. Nucl. Phys.} \textbf{\bibinfo{volume}{47}},
  \bibinfo{pages}{517} (\bibinfo{year}{2001}).

\bibitem[{\citenamefont{Fridmann et~al.}(2005)\citenamefont{Fridmann,
  Wiedenhover, Gade, Baby, Bazin, Brown, Campbell, Cook, Cottle, Diffenderfer
  et~al.}}]{ingo05}
\bibinfo{author}{\bibfnamefont{J.}~\bibnamefont{Fridmann}}
  {\it et~al.},
  \bibinfo{journal}{Nature} \textbf{\bibinfo{volume}{7044}},
  \bibinfo{pages}{922} (\bibinfo{year}{2005}).

\bibitem[{\citenamefont{Hansen and Sherrill}(2001)}]{hansen01}
\bibinfo{author}{\bibfnamefont{P.}~\bibnamefont{Hansen}} \bibnamefont{and}
  \bibinfo{author}{\bibfnamefont{B.}~\bibnamefont{Sherrill}},
  \bibinfo{journal}{Nucl. Phys.} \textbf{\bibinfo{volume}{A693}},
  \bibinfo{pages}{133} (\bibinfo{year}{2001}).

\bibitem[{\citenamefont{Caurier et~al.}(2005)\citenamefont{Caurier,
  Martinez-Pinedo, Nowacki, Poves, and Zuker}}]{caurier05}
\bibinfo{author}{\bibfnamefont{E.}~\bibnamefont{Caurier}},
  \bibinfo{author}{\bibfnamefont{G.}~\bibnamefont{Martinez-Pinedo}},
  \bibinfo{author}{\bibfnamefont{F.}~\bibnamefont{Nowacki}},
  \bibinfo{author}{\bibfnamefont{A.}~\bibnamefont{Poves}}, \bibnamefont{and}
  \bibinfo{author}{\bibfnamefont{A.~P.} \bibnamefont{Zuker}},
  \bibinfo{journal}{Rev. Mod. Phys.} \textbf{\bibinfo{volume}{77}},
  \bibinfo{pages}{427} (\bibinfo{year}{2005}).

\bibitem[{\citenamefont{Mahaux and Weidenm\"uller}(1969)}]{mahaux69}
\bibinfo{author}{\bibfnamefont{C.}~\bibnamefont{Mahaux}} \bibnamefont{and}
  \bibinfo{author}{\bibfnamefont{H.}~\bibnamefont{Weidenm\"uller}},
  \emph{\bibinfo{title}{Shell-Model Approach to Nuclear Reactions}}
  (\bibinfo{publisher}{North-Holland, Amsterdam},
  \bibinfo{year}{1969}).

\bibitem[{\citenamefont{Volya and
  Zelevinsky}(2003{\natexlab{a}})}]{volya_PRC67}
\bibinfo{author}{\bibfnamefont{A.}~\bibnamefont{Volya}} \bibnamefont{and}
  \bibinfo{author}{\bibfnamefont{V.}~\bibnamefont{Zelevinsky}},
  \bibinfo{journal}{Phys. Rev. C} \textbf{\bibinfo{volume}{67}},
  \bibinfo{pages}{54322} (\bibinfo{year}{2003}{\natexlab{a}}).

\bibitem[{\citenamefont{Volya and Zelevinsky}(2005)}]{volya_PRL}
\bibinfo{author}{\bibfnamefont{A.}~\bibnamefont{Volya}} \bibnamefont{and}
  \bibinfo{author}{\bibfnamefont{V.}~\bibnamefont{Zelevinsky}},
  \bibinfo{journal}{Phys. Rev. Lett.} \textbf{\bibinfo{volume}{94}},
  \bibinfo{pages}{052501} (\bibinfo{year}{2005}).

\bibitem[{\citenamefont{Feshbach}(1958)}]{feshbach58}
\bibinfo{author}{\bibfnamefont{H.}~\bibnamefont{Feshbach}},
  \bibinfo{journal}{Ann. Phys.} \textbf{\bibinfo{volume}{5}},
  \bibinfo{pages}{357} (\bibinfo{year}{1958}).

\bibitem[{\citenamefont{Feshbach}(1962)}]{feshbach62}
\bibinfo{author}{\bibfnamefont{H.}~\bibnamefont{Feshbach}},
  \bibinfo{journal}{Ann. Phys.} \textbf{\bibinfo{volume}{19}},
  \bibinfo{pages}{287} (\bibinfo{year}{1962}).

\bibitem[{\citenamefont{Rotter}(1991)}]{rotter91}
\bibinfo{author}{\bibfnamefont{I.}~\bibnamefont{Rotter}},
  \bibinfo{journal}{Rep. Prog. Phys.} \textbf{\bibinfo{volume}{54}},
  \bibinfo{pages}{635} (\bibinfo{year}{1991}).

\bibitem[{\citenamefont{Rotter}(2001)}]{rotter01}
\bibinfo{author}{\bibfnamefont{I.}~\bibnamefont{Rotter}},
  \bibinfo{journal}{Phys. Rev. E} \textbf{\bibinfo{volume}{64}},
  \bibinfo{pages}{036213} (\bibinfo{year}{2001}).

\bibitem[{\citenamefont{Weisskopf and Wigner}(1930)}]{weisskopf30}
\bibinfo{author}{\bibfnamefont{V.}~\bibnamefont{Weisskopf}} \bibnamefont{and}
  \bibinfo{author}{\bibfnamefont{E.}~\bibnamefont{Wigner}},
  \bibinfo{journal}{Z. Phys.} \textbf{\bibinfo{volume}{63}},
  \bibinfo{pages}{54} (\bibinfo{year}{1930}).

\bibitem[{\citenamefont{Rice}(1933)}]{rice33}
\bibinfo{author}{\bibfnamefont{O.}~\bibnamefont{Rice}}, \bibinfo{journal}{J.
  Chem. Phys.} \textbf{\bibinfo{volume}{1}}, \bibinfo{pages}{375}
  (\bibinfo{year}{1933}).

\bibitem[{\citenamefont{Fano}(1935)}]{fano35}
\bibinfo{author}{\bibfnamefont{U.}~\bibnamefont{Fano}}, \bibinfo{journal}{Nuovo
  Cim.} \textbf{\bibinfo{volume}{12}}, \bibinfo{pages}{156}
  (\bibinfo{year}{1935}).

\bibitem[{\citenamefont{Fano}(1961)}]{fano61}
\bibinfo{author}{\bibfnamefont{U.}~\bibnamefont{Fano}}, \bibinfo{journal}{Phys.
  Rev.} \textbf{\bibinfo{volume}{124}}, \bibinfo{pages}{1866}
  (\bibinfo{year}{1961}).

\bibitem[{\citenamefont{Betan et~al.}(2002)\citenamefont{Betan, Liotta,
  Sandulescu, and Vertse}}]{betan02}
\bibinfo{author}{\bibfnamefont{R.} \bibnamefont{Id Betan}},
  \bibinfo{author}{\bibfnamefont{R.~J.} \bibnamefont{Liotta}},
  \bibinfo{author}{\bibfnamefont{N.}~\bibnamefont{Sandulescu}},
  \bibnamefont{and} \bibinfo{author}{\bibfnamefont{T.}~\bibnamefont{Vertse}},
  \bibinfo{journal}{Phys. Rev. Lett.} \textbf{\bibinfo{volume}{89}},
  \bibinfo{pages}{042501} (\bibinfo{year}{2002}).

\bibitem[{\citenamefont{Michel et~al.}(2002)\citenamefont{Michel, Nazarewicz,
  Ploszajczak, and Bennaceur}}]{michel02}
\bibinfo{author}{\bibfnamefont{N.}~\bibnamefont{Michel}},
  \bibinfo{author}{\bibfnamefont{W.}~\bibnamefont{Nazarewicz}},
  \bibinfo{author}{\bibfnamefont{M.}~\bibnamefont{Ploszajczak}},
  \bibnamefont{and}
  \bibinfo{author}{\bibfnamefont{K.}~\bibnamefont{Bennaceur}},
  \bibinfo{journal}{Phys. Rev. Lett.} \textbf{\bibinfo{volume}{89}},
  \bibinfo{pages}{042502} (\bibinfo{year}{2002}).

\bibitem[{\citenamefont{Michel et~al.}(2003)\citenamefont{Michel, Nazarewicz,
  Ploszajczak, and Okolowicz}}]{michel03}
\bibinfo{author}{\bibfnamefont{N.}~\bibnamefont{Michel}},
  \bibinfo{author}{\bibfnamefont{W.}~\bibnamefont{Nazarewicz}},
  \bibinfo{author}{\bibfnamefont{M.}~\bibnamefont{Ploszajczak}},
  \bibnamefont{and}
  \bibinfo{author}{\bibfnamefont{J.}~\bibnamefont{Okolowicz}},
  \bibinfo{journal}{Phys. Rev. C} \textbf{\bibinfo{volume}{67}},
  \bibinfo{pages}{054311} (\bibinfo{year}{2003}).

\bibitem[{\citenamefont{Michel et~al.}(2004)\citenamefont{Michel, Nazarewicz,
  and Ploszajczak}}]{michel04}
\bibinfo{author}{\bibfnamefont{N.}~\bibnamefont{Michel}},
  \bibinfo{author}{\bibfnamefont{W.}~\bibnamefont{Nazarewicz}},
  \bibnamefont{and}
  \bibinfo{author}{\bibfnamefont{M.}~\bibnamefont{Ploszajczak}},
  \bibinfo{journal}{Phys. Rev. C} \textbf{\bibinfo{volume}{70}},
  \bibinfo{pages}{064313} (\bibinfo{year}{2004}).

\bibitem[{\citenamefont{Okolowicz et~al.}(2003)\citenamefont{Okolowicz,
  Ploszajczak, and Rotter}}]{okolowicz03}
\bibinfo{author}{\bibfnamefont{J.}~\bibnamefont{Okolowicz}},
  \bibinfo{author}{\bibfnamefont{M.}~\bibnamefont{Ploszajczak}},
  \bibnamefont{and} \bibinfo{author}{\bibfnamefont{I.}~\bibnamefont{Rotter}},
  \bibinfo{journal}{Phys. Rep.} \textbf{\bibinfo{volume}{374}},
  \bibinfo{pages}{271} (\bibinfo{year}{2003}).

\bibitem{hagen05} G. Hagen, M. Hjorth-Jensen, and J.S. Vaagen,
Phys. Rev. C {\bf 71}, 044314 (2005).
\bibitem{engelbrecht73} C.A. Engelbrecht and H.A.
Weidenm\"{u}ller, Phys. Rev. C {\bf 8}, 859 (1973).

\bibitem{durand76} L. Durand, Phys. Rev. D {\bf 14}, 3174 (1976).

\bibitem[{\citenamefont{Sokolov and Zelevinsky}(1989)}]{sokolov89}
\bibinfo{author}{\bibfnamefont{V.}~\bibnamefont{Sokolov}} \bibnamefont{and}
  \bibinfo{author}{\bibfnamefont{V.}~\bibnamefont{Zelevinsky}},
  \bibinfo{journal}{Nucl. Phys.} \textbf{\bibinfo{volume}{A504}},
  \bibinfo{pages}{562} (\bibinfo{year}{1989}).

\bibitem[{\citenamefont{Berggren}(1968)}]{berggren68}
\bibinfo{author}{\bibfnamefont{T.}~\bibnamefont{Berggren}},
  \bibinfo{journal}{Nucl. Phys.} \textbf{\bibinfo{volume}{A109}},
  \bibinfo{pages}{265} (\bibinfo{year}{1968}).

\bibitem[{\citenamefont{Siegert}(1939)}]{siegert39}
\bibinfo{author}{\bibfnamefont{A.~F.~J.} \bibnamefont{Siegert}},
  \bibinfo{journal}{Phys. Rev.} \textbf{\bibinfo{volume}{56}},
  \bibinfo{pages}{750} (\bibinfo{year}{1939}).

\bibitem[{\citenamefont{Breit and Wigner}(1936)}]{breit36}
\bibinfo{author}{\bibfnamefont{G.}~\bibnamefont{Breit}} \bibnamefont{and}
  \bibinfo{author}{\bibfnamefont{E.}~\bibnamefont{Wigner}},
  \bibinfo{journal}{Phys. Rev.} \textbf{\bibinfo{volume}{49}},
  \bibinfo{pages}{519} (\bibinfo{year}{1936}).

\bibitem[{\citenamefont{{Holt} et~al.}(1998)\citenamefont{{Holt}, {Engeland},
  {Hjorth-Jensen}, and {Osnes}}}]{sn132g}
\bibinfo{author}{\bibfnamefont{A.}~\bibnamefont{{Holt}}},
  \bibinfo{author}{\bibfnamefont{T.}~\bibnamefont{{Engeland}}},
  \bibinfo{author}{\bibfnamefont{M.}~\bibnamefont{{Hjorth-Jensen}}},
  \bibnamefont{and} \bibinfo{author}{\bibfnamefont{E.}~\bibnamefont{{Osnes}}},
  \bibinfo{journal}{Nucl. Phys.} \textbf{\bibinfo{volume}{A634}},
  \bibinfo{pages}{41} (\bibinfo{year}{1998}).

\bibitem[{\citenamefont{Brown and Wildenthal}(1988)}]{USD}
\bibinfo{author}{\bibfnamefont{B.A.}~\bibnamefont{Brown}} \bibnamefont{and}
  \bibinfo{author}{\bibfnamefont{B.}~\bibnamefont{Wildenthal}},
  \bibinfo{journal}{Ann. Rev. Nucl. Part. Sci.} \textbf{\bibinfo{volume}{38}},
  \bibinfo{pages}{29} (\bibinfo{year}{1988}).

\bibitem[{\citenamefont{Brown et~al.}(1994)\citenamefont{Brown, Etchegoyen, and
  Rae}}]{oxbash}
\bibinfo{author}{\bibfnamefont{B.A.}~\bibnamefont{Brown}},
  \bibinfo{author}{\bibfnamefont{A.}~\bibnamefont{Etchegoyen}},
  \bibnamefont{and} \bibinfo{author}{\bibfnamefont{W.}~\bibnamefont{Rae}},
  \bibinfo{type}{Tech. Rep.} \bibinfo{number}{MSU-NSCL 524},
  \bibinfo{institution}{NSCL, Michigan State University}
  (\bibinfo{year}{1994}).

\bibitem[{\citenamefont{Bohr and Motttelson}(1998)}]{BMI}
\bibinfo{author}{\bibfnamefont{A.}~\bibnamefont{Bohr}} \bibnamefont{and}
  \bibinfo{author}{\bibfnamefont{B.}~\bibnamefont{Motttelson}},
  \emph{\bibinfo{title}{Nuclear Structure}}, vol I (\bibinfo{publisher}{World
  Scientific Publishing}, \bibinfo{year}{1998}).

\bibitem[{\citenamefont{Landau and Lifshitz}(1981)}]{landauIII}
\bibinfo{author}{\bibfnamefont{L.}~\bibnamefont{Landau}} \bibnamefont{and}
  \bibinfo{author}{\bibfnamefont{E.}~\bibnamefont{Lifshitz}},
  \emph{\bibinfo{title}{Quantum Mechanics, Non-relativistic Theory.}} Third
  edition (\bibinfo{publisher}{Pergamon Press, New
  York}, \bibinfo{year}{1981}).

\bibitem[{\citenamefont{Merzbacher}(1998)}]{merzbacher}
\bibinfo{author}{\bibfnamefont{E.}~\bibnamefont{Merzbacher}},
  \emph{\bibinfo{title}{Quantum Mechanics}} (\bibinfo{publisher}{John Wiley and Sons, inc.},
  \bibinfo{address}{New York}, \bibinfo{year}{1998}).

\bibitem[{\citenamefont{Davids and Esbensen}(2000)}]{davids00}
\bibinfo{author}{\bibfnamefont{C.}~\bibnamefont{Davids}} \bibnamefont{and}
  \bibinfo{author}{\bibfnamefont{H.}~\bibnamefont{Esbensen}},
  \bibinfo{journal}{Phys. Rev. C} \textbf{\bibinfo{volume}{61}},
  \bibinfo{pages}{054302} (\bibinfo{year}{2000}).

\bibitem[{\citenamefont{Burgov and Kadmensky}(1989)}]{burgov89}
\bibinfo{author}{\bibfnamefont{V.}~\bibnamefont{Burgov}} \bibnamefont{and}
  \bibinfo{author}{\bibfnamefont{S.}~\bibnamefont{Kadmensky}},
  \bibinfo{journal}{Sov. J. Nucl. Phys.} \textbf{\bibinfo{volume}{49}},
  \bibinfo{pages}{967} (\bibinfo{year}{1989}).

\bibitem[{\citenamefont{Burgov and Kadmensky}(1996)}]{burgov96}
\bibinfo{author}{\bibfnamefont{V.}~\bibnamefont{Burgov}} \bibnamefont{and}
  \bibinfo{author}{\bibfnamefont{S.}~\bibnamefont{Kadmensky}},
  \bibinfo{journal}{Phys. At. Nucl.} \textbf{\bibinfo{volume}{59}},
  \bibinfo{pages}{424} (\bibinfo{year}{1996}).

\bibitem[{\citenamefont{Wigner}(1948)}]{wigner48}
\bibinfo{author}{\bibfnamefont{E.}~\bibnamefont{Wigner}},
  \bibinfo{journal}{Phys. Rev.} \textbf{\bibinfo{volume}{73}},
  \bibinfo{pages}{1002} (\bibinfo{year}{1948}).

\bibitem[{\citenamefont{Baz et~al.}(1971)\citenamefont{Baz, Zeldovich, and
  Perelomov}}]{baz71}
\bibinfo{author}{\bibfnamefont{A.}~\bibnamefont{Baz}},
  \bibinfo{author}{\bibfnamefont{I.}~\bibnamefont{Zeldovich}},
  \bibnamefont{and}
  \bibinfo{author}{\bibfnamefont{A.}~\bibnamefont{Perelomov}},
  \emph{\bibinfo{title}{Scattering, reactions and decay in nonrelativistic
  quantum mechanics}} (\bibinfo{publisher}{Israel Program for Scientific Translations, Jerusalem},
  \bibinfo{year}{1969}).

\bibitem{VZCovello02} V. Zelevinsky and A. Volya, in {\sl
Challenges of Nuclear Structure}, ed. A. Covello (World
Scientific, Singapore, 2002) p. 261.

\bibitem[{\citenamefont{Dicke}(1954)}]{dicke54}
\bibinfo{author}{\bibfnamefont{R.}~\bibnamefont{Dicke}},
  \bibinfo{journal}{Phys. Rev.} \textbf{\bibinfo{volume}{93}},
  \bibinfo{pages}{99} (\bibinfo{year}{1954}).

\bibitem{rotureau05}J. Rotureau, J. Okolowicz, and M. Ploszajczak, Phys. Rev.
Lett. {\bf 95}, 042503 (2005).
\bibitem{brown03} B.A. Brown, and F.C. Barker, Phys. Rev. C {\bf 67}, 041304 (2003).
\bibitem{grigorenko03}L.V. Grigorenko and M.V. Zhukov, Phys. Rev. C {\bf 68} 054005 (2003).
\bibitem{barker99} F. C. Barker, Phys. Rev. C {\bf 59}, 353 (1999).
\bibitem[{\citenamefont{Cohen and Kurath}(1965)}]{cohen65}
\bibinfo{author}{\bibfnamefont{S.}~\bibnamefont{Cohen}} \bibnamefont{and}
  \bibinfo{author}{\bibfnamefont{D.}~\bibnamefont{Kurath}},
  \bibinfo{journal}{Nucl. Phys.} \textbf{\bibinfo{volume}{A73}},
  \bibinfo{pages}{1} (\bibinfo{year}{1965}).

\bibitem[{\citenamefont{Stevenson et~al.}(1988)\citenamefont{Stevenson, Brown,
  Chen, Clayton, Kashy, Mikolas, Nolen, Samuel, Sherrill, Winfield
  et~al.}}]{stevenson88}
\bibinfo{author}{\bibfnamefont{J.}~\bibnamefont{Stevenson}}
  {\it et~al.}, \bibinfo{journal}{Phys. Rev. C}
  \textbf{\bibinfo{volume}{37}}, \bibinfo{pages}{2220} (\bibinfo{year}{1988}).

\bibitem[{\citenamefont{J\"anecke and {\sl et~al.}}(1996)}]{janecke96}
\bibinfo{author}{\bibfnamefont{J.}~\bibnamefont{J\"anecke}}
  \bibinfo{author}{\bibnamefont{{\it et~al.}}}, \bibinfo{journal}{Phys. Rev.
  C} \textbf{\bibinfo{volume}{54}}, \bibinfo{pages}{1070}
  (\bibinfo{year}{1996}).

\bibitem[{\citenamefont{Nakayama and {\sl et. al.}}(2000)}]{nakayama00}
\bibinfo{author}{\bibfnamefont{S.}~\bibnamefont{Nakayama}}
  \bibinfo{author}{\bibnamefont{{\it et~al.}}}, \bibinfo{journal}{Phys. Rev.
  Lett.} \textbf{\bibinfo{volume}{85}}, \bibinfo{pages}{262}
  (\bibinfo{year}{2000}).

\bibitem[{nnd()}]{nndc}
\emph{\bibinfo{title}{Evaluated nuclear structure data file}},
  \bibinfo{note}{http://www.nndc.bnl.gov}.

\bibitem[{\citenamefont{Rogachev et~al.}(2004)\citenamefont{Rogachev,
  Boutachkov, Aprahamian, Becchetti, Bychowski, Chen, Chubarian, DeYoung,
  Goldberg, Kolata et~al.}}]{rogachev04}
\bibinfo{author}{\bibfnamefont{G.~V.} \bibnamefont{Rogachev}}
  {\it et~al.}, \bibinfo{journal}{Phys. Rev. Lett.}
  \textbf{\bibinfo{volume}{92}}, \bibinfo{eid}{232502}
  (\bibinfo{year}{2004}).

\bibitem[{\citenamefont{Korsheninnikov et~al.}(1999)}]{korsheninnikov99}
\bibinfo{author}{\bibfnamefont{A.~A.} \bibnamefont{Korsheninnikov}}
  {\it et~al.}, \bibinfo{journal}{Phys. Rev. Lett.}
  \textbf{\bibinfo{volume}{82}}, \bibinfo{pages}{3581} (\bibinfo{year}{1999}).

\bibitem[{\citenamefont{Auerbach et~al.}(2004)\citenamefont{Auerbach,
  Zelevinsky, and Volya}}]{pentaquark}
\bibinfo{author}{\bibfnamefont{N.}~\bibnamefont{Auerbach}},
  \bibinfo{author}{\bibfnamefont{V.}~\bibnamefont{Zelevinsky}},
  \bibnamefont{and} \bibinfo{author}{\bibfnamefont{A.}~\bibnamefont{Volya}},
  \bibinfo{journal}{Phys. Lett. B} \textbf{\bibinfo{volume}{590}},
  \bibinfo{pages}{45} (\bibinfo{year}{2004}).

\bibitem[{\citenamefont{Volya and Zelevinsky}(2003{\natexlab{b}})}]{JOB5}
\bibinfo{author}{\bibfnamefont{A.}~\bibnamefont{Volya}} \bibnamefont{and}
  \bibinfo{author}{\bibfnamefont{V.}~\bibnamefont{Zelevinsky}},
  \bibinfo{journal}{J. Opt. B} \textbf{\bibinfo{volume}{5}},
  \bibinfo{pages}{S450} (\bibinfo{year}{2003}{\natexlab{b}}).

\bibitem[{\citenamefont{Teichmann}(1050)}]{teichmann50}
\bibinfo{author}{\bibfnamefont{T.}~\bibnamefont{Teichmann}},
  \bibinfo{journal}{Phys. Rev.} \textbf{\bibinfo{volume}{77}},
  \bibinfo{pages}{506} (\bibinfo{year}{1050}).

\bibitem[{\citenamefont{Teichmann and Wigner}(1952)}]{teichmann52}
\bibinfo{author}{\bibfnamefont{T.}~\bibnamefont{Teichmann}} \bibnamefont{and}
  \bibinfo{author}{\bibfnamefont{E.}~\bibnamefont{Wigner}},
  \bibinfo{journal}{Phys. Rev.} \textbf{\bibinfo{volume}{87}},
  \bibinfo{pages}{123} (\bibinfo{year}{1952}).

\bibitem[{\citenamefont{Brown et~al.}(1988)\citenamefont{Brown, Richter,
  Julies, and Wildenthal}}]{HBUSD}
\bibinfo{author}{\bibfnamefont{B.A.}~\bibnamefont{Brown}},
  \bibinfo{author}{\bibfnamefont{W.}~\bibnamefont{Richter}},
  \bibinfo{author}{\bibfnamefont{R.}~\bibnamefont{Julies}}, \bibnamefont{and}
  \bibinfo{author}{\bibfnamefont{B.}~\bibnamefont{Wildenthal}},
  \bibinfo{journal}{Ann. Phys. (N.Y.)} \textbf{\bibinfo{volume}{182}},
  \bibinfo{pages}{191} (\bibinfo{year}{1988}).

\bibitem[{\citenamefont{Skorodumov et~al.}(2005)\citenamefont{Skorodumov,
  Rogachev, Poutachkov, Aprahamian, Kolta, Lamm, Quinn, and
  Woehr}}]{skorodumov05}
\bibinfo{author}{\bibfnamefont{B.}~\bibnamefont{Skorodumov}}
{\sl et~al.},
  \bibinfo{journal}{submitted for publication to Phys. Rev. C.}
  (\bibinfo{year}{2005}).

\bibitem[{\citenamefont{Goldberg et~al.}(2004)\citenamefont{Goldberg,
  Chubarian, Tabacaru, Trache, Tribble, Aprahamian, Rogachev, Skorodumov, and
  Tang}}]{goldberg04}
\bibinfo{author}{\bibfnamefont{V.}~\bibnamefont{Goldberg}}
{\it et~al.},
  \bibinfo{journal}{Phys. Rev. C} \textbf{\bibinfo{volume}{69}},
  \bibinfo{pages}{31302} (\bibinfo{year}{2004}).

\bibitem[{\citenamefont{Tilley et~al.}()\citenamefont{Tilley, Cheves, Kelley,
  Raman, Weller, and Ajzenberg-Selove}}]{tunl}
\bibinfo{author}{\bibfnamefont{D.}~\bibnamefont{Tilley}},
  \bibinfo{author}{\bibfnamefont{C.}~\bibnamefont{Cheves}},
  \bibinfo{author}{\bibfnamefont{J.}~\bibnamefont{Kelley}},
  \bibinfo{author}{\bibfnamefont{S.}~\bibnamefont{Raman}},
  \bibinfo{author}{\bibfnamefont{H.}~\bibnamefont{Weller}}, \bibnamefont{and}
  \bibinfo{author}{\bibfnamefont{F.}~\bibnamefont{Ajzenberg-Selove}},
  \emph{\bibinfo{title}{Energy Levels of Light Nuclei, A = 3 - 20}},
  \bibinfo{note}{http://www.tunl.duke.edu/nucldata}.

\bibitem{bair66} J.K. Bair, J.L.C. Ford, Jr., and C.M. Jones, Phys. Rev.
{\bf 144}, 799 (1966).
\bibitem[{\citenamefont{S.F. et~al.}(1981)\citenamefont{S.F., Kinsey, and
  Dunford}}]{neutron}
\bibinfo{author}{\bibfnamefont{S.~F.} \bibnamefont{Mughabghab}},
  \bibinfo{author}{\bibfnamefont{R.}~\bibnamefont{Kinsey}}, \bibnamefont{and}
  \bibinfo{author}{\bibfnamefont{C.}~\bibnamefont{Dunford}},
  \emph{\bibinfo{title}{Neutron Cross Sections Series}}
  (\bibinfo{publisher}{Academic Press, New York}, \bibinfo{year}{1981}).

\bibitem[{\citenamefont{McLane et~al.}(1988)\citenamefont{McLane, Dunford, and
  Rose}}]{neutron2}
\bibinfo{author}{\bibfnamefont{V.}~\bibnamefont{McLane}},
  \bibinfo{author}{\bibfnamefont{C.}~\bibnamefont{Dunford}}, \bibnamefont{and}
  \bibinfo{author}{\bibfnamefont{P.}~\bibnamefont{Rose}},
  \emph{\bibinfo{title}{Neutron Cross Section Curves}},
  vol.~\bibinfo{volume}{2} (\bibinfo{publisher}{Academic Press, Boston},
  \bibinfo{year}{1988}).

\end{thebibliography}

\end{document}